\documentclass[twocolumn]{aastex63}
\usepackage[cmex10]{amsmath}
\usepackage{amssymb}	
\received{??}
\revised{??}
\accepted{?? draft \today}
\submitjournal{PASP}
\shorttitle{Interferometric Antenna Corrections}
\shortauthors{Cotton \& Mauch}

\begin{document}

\title{Correction of Radio Interferometric Imaging for Antenna Patterns}

\correspondingauthor{William Cotton}
\email{bcotton@nrao.edu}

\author{W.~D.~Cotton}
\affiliation{National Radio Astronomy Observatory \\
520 Edgemont Road \\
Charlottesville, VA 22903, USA}

\author{T.~Mauch}
\affiliation{South African Radio Astronomy Observatory, 2 Fir St.,
  Observatory, South Africa}  



\begin{abstract}
We describe and demonstrate a technique for correcting direction
dependent artifacts due to asymmetries in antenna patterns and
differences among antennas used in radio interoferometric imaging.
The technique can correct images in all Stokes parameters I, Q, U and V
and is shown with simulated data to reduce the level of artifacts to
near the level of those from the basic imaging technique.
The demonstrations use simulations of a mixed array of 13.5 and 15 m
antennas with asymmetric patterns.
The flux densities and spectral indices of the sources in a high dynamic
range realistic simulated sky model are well recovered.
Source polarization properties are also recovered in tests using
unpolarized and partly polarized sources.
The additional computational run time for Stokes I correction is about
50\% in a realistic test described. 
\end{abstract}



\section{Introduction} 
Radio interferometric arrays produce images of the sky as modified by
the beam pattern of the individual antennas.
Asymmetries in 2 axis alt-az mounted antennas introduce time dependent
gain variations in the direction of sources off the axis as the
antenna pattern rotates on the sky with parallactic
angle\footnote{ASKAP has a 3 axis mount to eliminate this effect
\citep{ASKAP}.}.  
This is the equivalent of observing a time varying source and will
introduce non-convolutional artifacts in the derived image.
The use of heterogeneous arrays, arrays of antennas with dissimilar
patterns on the sky, are an even more extreme case as the different
antennas see a different effective sky even at the same time.
Artifacts from bright sources caused by beam asymmetries and
differences among antennas can seriously corrupt the imaging of weaker
sources. 
It is worth noting that accurate correction for beam models requires
accurate pointing of the antennas. 

Antenna design can optimize polarization performance over only a
limited area of the antenna beam on the sky.
Away from this limited area, the antennas introduce a spurious
polarized response to total intensity.
As source fractional polarization can be a few percent or less, this
instrumental response can dwarf source polarization in regions of the
antenna pattern with comparable, or larger, instrumental
polarizations.
Observations over a range of parallactic angle (observing geometry)
tends to reduce, but not eliminate this effect.
This effect can render polarization measurements of limited use over a
significant portion of the, otherwise useful, antenna beam.
An array with heterogeneous antenna designs further complicates this
problem. 

This paper describes a technique for correcting derived images for the
effects of antenna patterns and presents tests using simulations of
the heterogeneous array MeerKAT+ (13.5 m MeerKAT dishes plus 15 m SKA
dishes) as implemented in the Obit package
\citep{OBIT}\footnote{http://www.cv.nrao.edu/$\sim$bcotton/Obit.html}.    

\section{Direction Dependent Effects}
Direction dependent effects (DDEs) have long been recognized as a
fundamental limitation to radio interferometric imaging.
Direction dependent effects are always present in the form of the
power gain pattern of the interferometer elements but in cases in
which they are constant during the observations they can be dealt with 
using a ``Primary beam correction'' of the final image.
This can be the case when the elements have an equatorial mount or a
third axis ``field rotator'' as is used in ASKAP \citep{ASKAP}.

Time variable direction dependent effects such as antenna pointing
errors, circular asymmetries in alt-az mounted antennas, variations
in atmospheric phase across the field of view and the like will
introduce image artifacts which will decrease the dynamic range and
may obscure fainter objects in the presence of brighter ones.
Mitigation of these are needed if they limit the science.
A general discussion of DDEs with a formalism to describe them is
given in \cite{Smirnov2011}

DDEs come in two basic flavors, those which must be inferred from the
observations themselves and those which can be derived from models of
the instrument or external measurements of the atmosphere.
Techniques which derive the corrections from the data can, and have,
been used to correct effects that could be modeled.
Data sets contain a finite amount of information; calibration inferred
from the data will use some of these ``degrees of freedom'' that could
otherwise be used to improve the image quality. 
Correction based on external measurements is thus preferred to that
derived from the data.
In practice, both predictable and unpredictable effects will be
present in most data.

\subsection{DDEs inferred from the observations}
Most attempts to date to correct DDEs have, of necessity, inferred the
effects from the observations.
\subsubsection{Peeling}
The flux density distribution of radio sources is such that at lower
frequencies, artifacts are frequently dominated by a few bright
sources.
These can be frequently greatly reduced using ``Peeling''
\citep{Noordam2004} in which the complex antenna gains in the direction
of the bright sources are derived and the contribution of the peeled
source subtracted from the data set.
If multiple sources need peeling they are processed sequentially.
An example of peeling is given in \cite{ObitMemo72}.
\subsubsection{Differential gains}
A more sophisticated approach to peeling multiple sources is the
``differential gain'' method of \cite{Smirnov2011}.
Complex gains in the directions of multiple sources are solved for
simultaneously. 
This reduces the buildup of numerical errors from sequentially peeling
multiple sources.
\subsubsection{Clustered calibration}
Especially at low frequencies where the ionosphere is important,
deriving the gains in the directions of a few bright sources may be
inadequate. 
On the other hand, most sources are too weak to be individually
detectable.
The ``Clustered calibration'' described by \cite{Kazemi2013}
divides the sky up into regions with sufficient number of sources that
a set of collective gains can be derived.
These gains are then used to calibrate and image the respective regions.
\subsubsection{Field based calibration}
A different approach to ionospheric calibration is ``Field based
calibration'' of \citep{Cotton2004,Cotton2005} which is
applicable in the regime \citep{Lonsdale2005} in which the geometry of
the field of view is distorted but not on the scale of individual
sources.
Snapshot imaging of strong sources in the field allows time sampling of
this distortion field which is fitted with a Zernike phase screen.
This screen is used to dedistort/redistort the data during deconvolution. 
This technique has been applied to VLA data at 74 \citep{Lane2014} and
327 \citep{Uson2012} MHz.
\subsubsection{SPAM}
A generalization of Field based calibration is the ``SPAM'' (Source
Peeling And Modeling) of \cite{Intema2009} which fits a function to
the phases derived from peeling the bright sources in the field.
This should work even when the individual sources start becoming
distorted. 
\subsubsection{Facet Calibration}
A refinement of the Clustered calibration is the ``Facet Calibration''
of \cite{vanWeeren2016}.
The initial direction independent calibration is used to subtract all
sources in the initial CLEAN from the data and the field of view is
divided into a compact set of facets. 
Sources from a given, brightest facet are added back and a
single set of calibration parameters for that facet is derived.
The facet's sources are then subtracted, correcting for these DDEs
before moving to the next brightest facet.

\subsection{Modeled DDEs\label{previous}}
At present, the DDEs that can be effectively modeled are the antenna
gains which are modeled either by EM calculations or holography.
This has been done for MeerKAT \cite{Asad2021} and the VLA
\cite{Iheanet2019} at L band.
\subsubsection{AW Projection}
Antenna gains are a multiplicative effect in the image domain hence
are convolutional in the Fourier domain.
The ``AW projection'' adaptation of ``W projection''
\citep{A_projection} by \cite{Bhatnagar2013} allows including
corrections for time and direction dependent antenna gains
in the convolution kernel used in gridding the data.
These gains must then be unapplied in the degridding step when
subtracting the sky model from the visibility data.
An application to LOFAR is described in \cite{Tasse2013}.

When ionospheric phase fluctuations are rapid, full AW projection gets
very computationally expensive.
\cite{vanderTol2018} describes a technique for reducing this cost.

\subsubsection{DDFacet}
An alternate approach to reducing the cost of AW projection is using
image plane facets as described in \cite{Tasse2018} (DDFacet).
DDFacet can apply full Jones externally defined models of antenna beam
or other instrumental or atmospheric transmission effects.
Corrections applied in imaging must be unapplied in the degridding
step.
An application of DDFacet to deep LOFAR imaging is given in
\cite{Tasse2021}. 

\section{Antenna Beam Corrections \label{Beam}}
The technique developed in the following is to ignore the effects of
antenna beam asymmetries and differences among antennas when making
dirty/residual images in a CLEAN based deconvolution but then
calculating and subtracting an accurate instrumental response to a
partial sky model.
This calculation uses the approximation that the sky model only
includes the Stokes parameter being deconvolved.

In order to avoid numerical problems in regions of the image where the
amplitude of the parallel hand antenna pattern is low, the ratio of
the true antenna pattern to a ``perfect'' well behaved, symmetric,
real beam is used to correct the instrumental response to the CLEAN
sky model. 
After multiple major cycles, the residuals approach zero and
the sky model approaches the true sky model modified by the
``perfect'' antenna pattern.
Each major cycle needs to be fairly shallow to avoid incorporating
artifacts into the CLEAN sky model.

This procedure can correct the spurious off-axis response to any I, Q,
U or V but artifacts arising from Stokes I can seriously adversely
affect the images of polarized emission so it should be done first.
The final Stokes I sky model can then be used to subtract the array's
response from the initial dataset which is then used to image Q, U and
V.
This procedure is then repeated for Stokes Q, U and V, subtracting the
response to the final CLEAN model before moving on to the next Stokes
parameter. 

We adopt the formalism of \cite{smirnov11} to describe the direction
dependent response of an interferometer.
Assuming that the usual direction independent calibrations have been
applied to the data and there are no direction dependent effects other
than the beam shapes, the instrumental response for baseline $a-b$ in
terms of 2$\times$2 complex matrices becomes \citep{smirnov11}
\begin{equation}
V_{ab}(\nu,t)\ =\ \int_{l}\int_{m} E'_{a,\nu,t,l,m} X_{\nu,t,l,m}
E'^\dagger_{b,\nu,t,l,m} dl dm/n,
\label{InstResponse}
\end{equation}
where $l$ and $m$ (with $n\ =\ \sqrt{1-l^2-m^2}$ ) are direction
cosines; ${E'}_{a,\nu,t,l,m}$ and ${E'}_{b,\nu,t,l,m}$ are the ratios of the complex
antenna beam responses in the direction of $(l,m)$ at time $t$
and frequency $\nu$ for the first and second antennas of the baseline
to the ``perfect'' antenna beam pattern and $^\dagger$ signifies the
conjugate transpose.  
$X_{\nu,t,l,m}$ is given by
$$
X_{\nu,t,l,m}\ =\ B_{\nu,l,m}
  e^{-2\pi (u_{\nu,t} l + v_{\nu,t} m + w_{\nu,t} (n-1))}.
$$
$B_{\nu,l,m}$ is the ``source coherence matrix''and
$u_{\nu,t},v_{\nu,t},w_{\nu,t}$ are the spatial frequency coordinates
of baseline $a-b$ at time $t$ and frequency $\nu$.

Replacing the continuous sky brightness with $N$ discrete sources
Equation \ref{InstResponse} becomes
\begin{equation}
V_{ab}(\nu,t)\ =\ \sum_{j=1}^{N} E'_{a,\nu,t,l_j,m_j} X_{\nu,t,l_j,m_j}
E'^\dagger_{b,\nu,t,l_j,m_j}.
\label{InstResponseDis}
\end{equation}

The  ratio of the true beam pattern for antenna $k$ (a or b) to the
``perfect'' beam corresponding to the position of component j at
frequency $\nu$ and time $t$ can be expressed as
\begin{multline}
{E'}_{k,j,\nu,t}\ =
  \begin{bmatrix}
    pp^{ratio}_{k,j,\nu,t} & pq^{ratio}_{k,j,\nu,t} \\
    qp^{ratio}_{k,j,\nu,t}  & qq^{ratio}_{k,j,\nu,t,j}
  \end{bmatrix}
\label{ResponseModel}
\end{multline}
where $pp^{ratio}$,$pq^{ratio}$,$qp^{ratio}$, and $qq^{ratio}$ are the
ratios of the combinations of the two orthogonally polarized feeds. 
The variation in time is due to the parallactic angle change of alt-az
mounted antennas tracking the source.

The source coherency matrix for component $j$, $B^j$,  for baseline
$a-b$ at time $t$ and frequency 
$\nu$ for circular basis feeds is given by
\begin{equation}
 B^j_{a-b,\nu,t}\ = {1\over{2}}
  \begin{bmatrix}
    I_\nu^j+V_\nu^j & Q_\nu^j+iU_\nu^j \\
    Q_\nu^j+iU_\nu^j & I_\nu^j-V_\nu^j
  \end{bmatrix}
\label{VisModelCir}
\end{equation}
where $I_\nu^j$, $Q_\nu^j$, $U_\nu^j$ and $V_\nu^j$ are the Stokes
parameters of component $j$ and $i$ is $\sqrt{-1}$ and for linear
feeds is given by 
\begin{equation}
 B^j_{a-b,\nu,t}\ = {1\over{2}} \\
  \begin{bmatrix}
    I^j_\nu+Q'^j_\nu & U'^j_\nu + iV^j_\nu\\
    U'^j_\nu -iV^j_\nu & I^j_\nu - Q'^j_\nu)
  \end{bmatrix}
\label{VisModelLin}
\end{equation}
where
$$Q'^j\ = Q^j_\nu cos(2\psi) + U^j_\nu sin(2\psi),$$
$$U'^j\ = -Q^j_\nu sin(2\psi) + U^j_\nu cos(2\psi)$$
and the parallactic angle $\psi$ is 
\begin{equation}\label{ParAng}
\psi\ =\ {\rm tan}^{-1}\Big({{{\rm cos}\ \lambda\ {\rm sin}\ h}
  \over{{\rm sin}\ \lambda\ {\rm cos}\ \delta\ -\ {\rm cos}\ \lambda\
    {\rm sin}\ \delta\ {\rm cos}\ h}}
\Big).
\end{equation}

Subtracting $V(\nu,t)$ (eq. \ref{InstResponseDis}) from the observed data will
remove the response to the sky model including the spurious
instrumental polarization terms.
In a wide variety of cases, Stokes I dominates Q and U which in turn
are much stronger than Stokes V.
In this case Stokes I can be imaged and deconvolved with the
approximation that Q,U and V are zero.
Once the response to Stokes I is removed, Q and U can be deconvolved
with the approximation that I and V are zero.

\section{Simulations\label{Simulations}}
\begin{figure} 
    \includegraphics[width=\columnwidth]{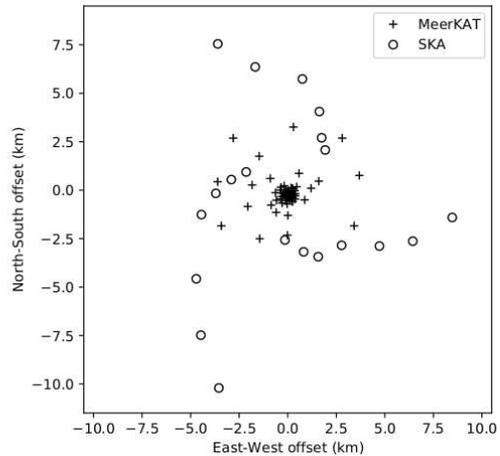} 
    \caption{The MeerKAT+ array configuration used in the simulation. Positions
             of antennas are shown in km relative to the array center at
             latitude $-$30:42:39.8, longitude +21:26:38.0. The positions of
             the 64 13.5\,m diameter MeerKAT dishes are shown as plus signs and
             the 20 15.0\,m SKA dishes are shown as circles.}
    \label{arr_config}
\end{figure}

\begin{figure*}
    \includegraphics[width=0.5\textwidth]{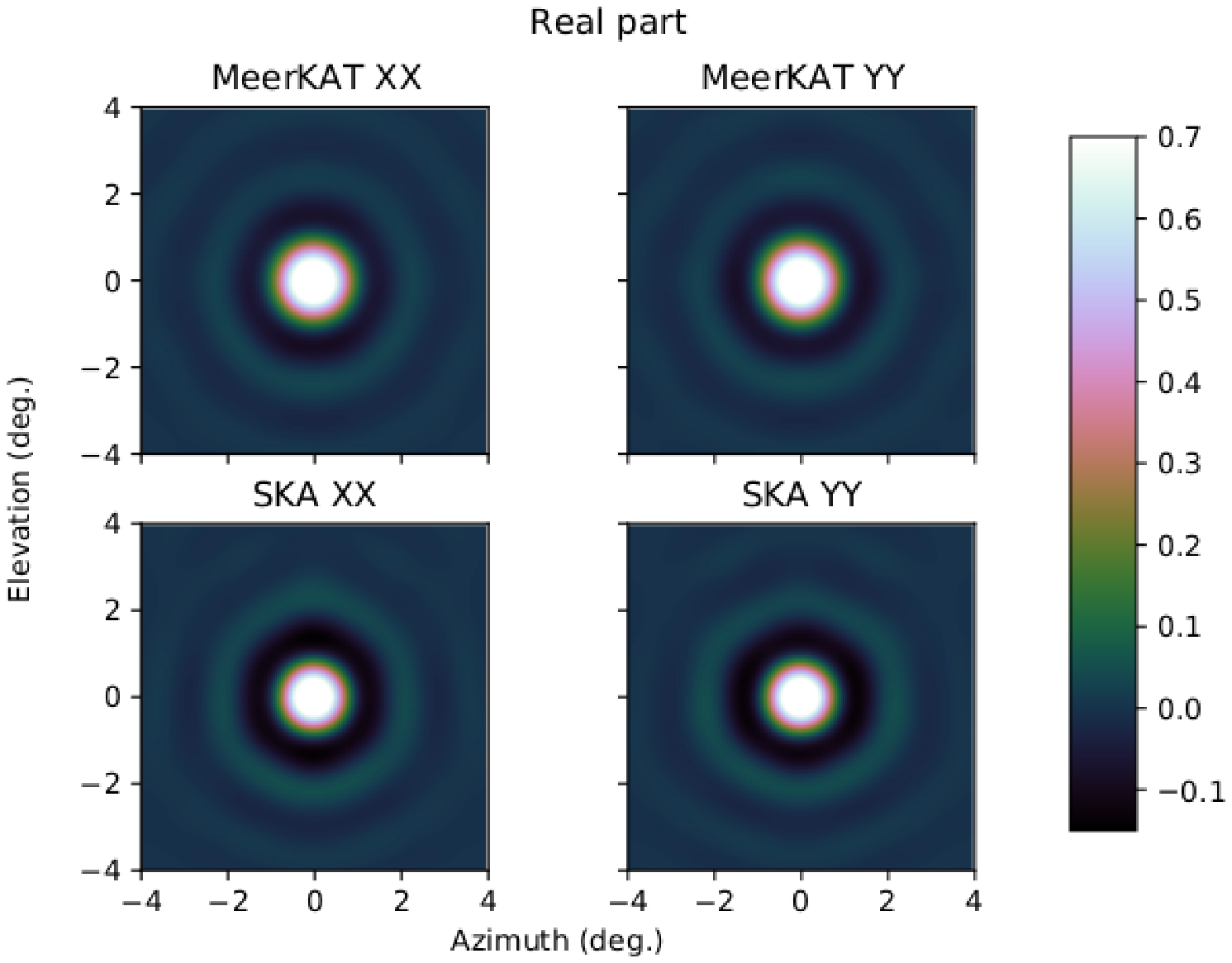}%
    \includegraphics[width=0.5\textwidth]{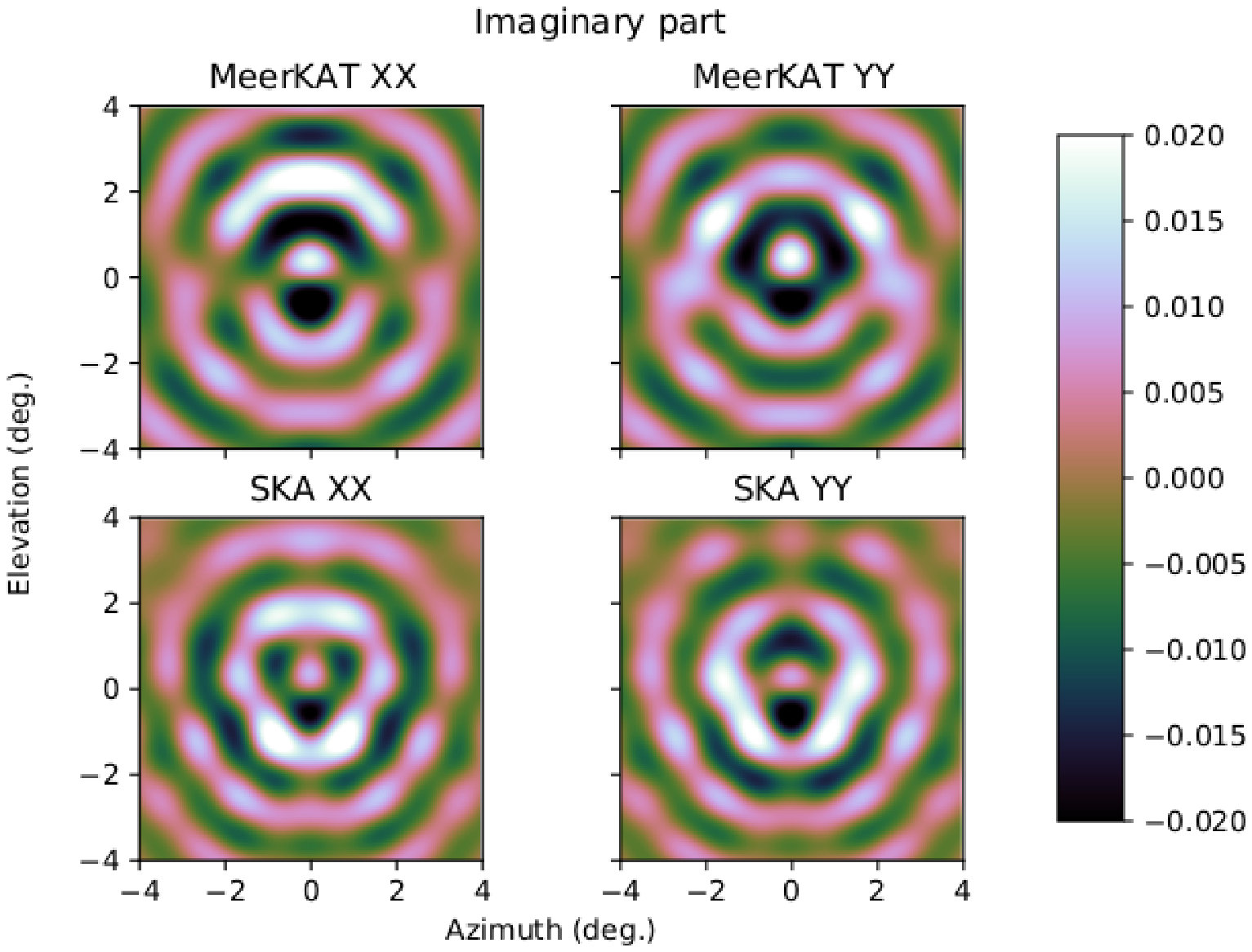}
    \includegraphics[width=0.5\textwidth]{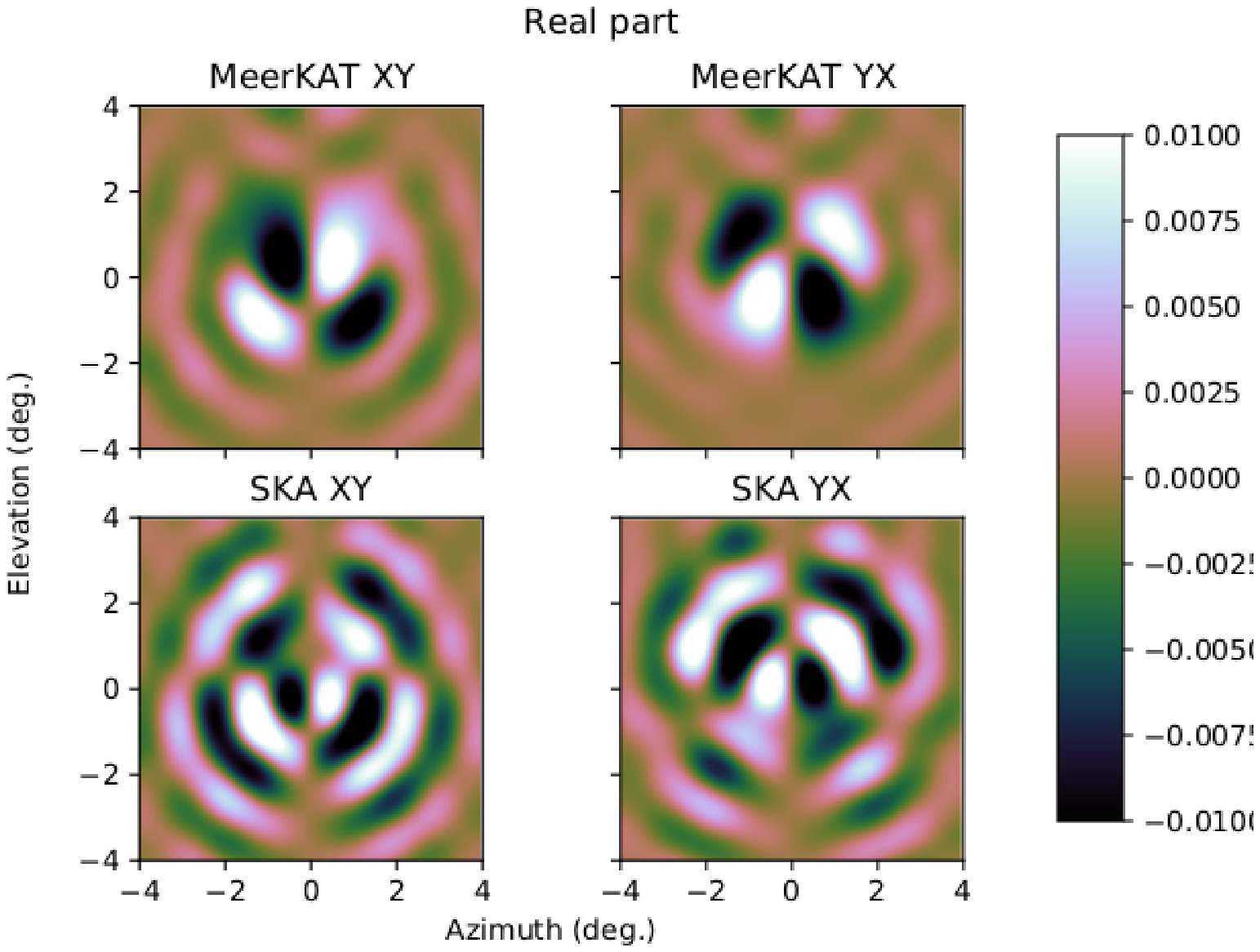}%
    \includegraphics[width=0.5\textwidth]{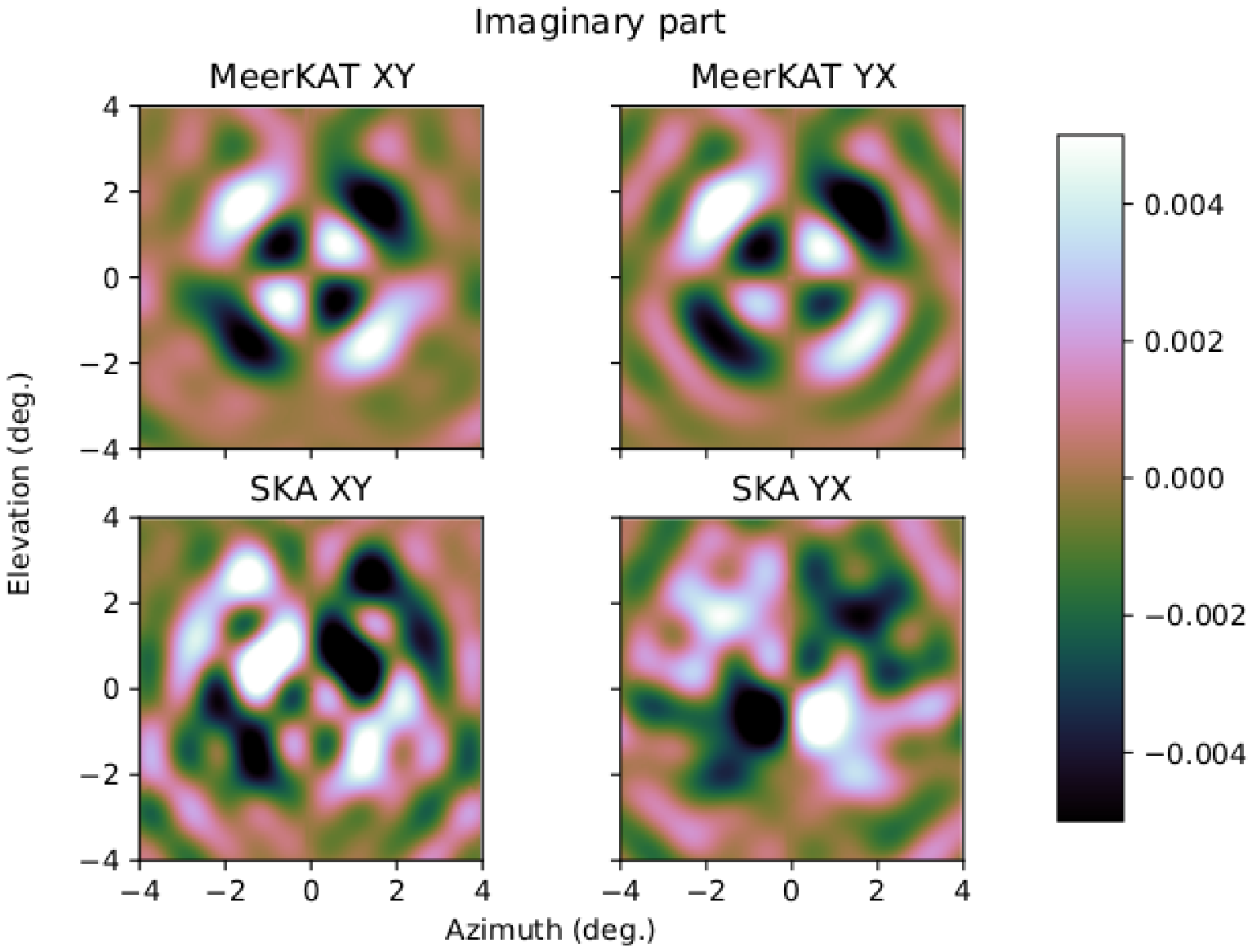}
    \caption{Real (left) and imaginary (right) parts of the complex voltage
             patterns at 1510\,MHz used in the simulation for 13.5\,m MeerKAT
             dishes and 15.0\,m SKA dishes (as labeled). The co-polarization
             response is shown in the top two rows and the cross-polarization
             in the bottom two.}
    \label{patt_volt}
\end{figure*}

Simulations use the planned MeerKAT+ array  which will include
antennas located out to distances of nearly 10\,km from its
center at latitude $-$30:42:39.8, longitude +21:26:38.0. 
The array will be made up of a `core' consisting of the original
MeerKAT 13.5\,m dishes whose locations are shown as plus signs in
Figure~\ref{arr_config}. 
These will be supplemented by further dishes primarily at greater
distance from the array center, these will have a 15.0\,m diameter SKA
design (circles in Figure~\ref{arr_config}). 
The different apertures of the MeerKAT and SKA dishes result in them
having significantly different attenuation patterns. 
All antennas are equipped with feeds sensitive to linear polarization.
Complex voltage patterns at 1510\,MHz derived from simulations of the
MeerKAT antenna optics made by the EMSS \citep{Asad2021} were provided by
M.~De~Villiers and are shown in Figure~\ref{patt_volt}.

The simulations described in this paper all use a pointing center at
$\alpha = 0^h$, $\delta = -80^\circ$, with a single `track' of 120 time samples
over a 12 hour observation period starting at 1970-01-01T15:50:15 UTC (when
the array pointing has parallactic angle $\psi\sim0^\circ$). The MeerKAT+ frequency
range at L-band was sampled with 64 channels over the range 898.8--1669.2\,MHz
and the 84 antennas with positions and types as shown in
Figure~\ref{arr_config}.

\subsection{Simulated visibilities}
\label{sim_vis}

The simulation defines a function that computes complex visibilities given an
observational setup and a list of point sources to be simulated. The
visibilities are simulated using the M\"{u}ller and Jones matrix
formalism described by \cite{smirnov11} and references therein.

The flux density of any point source at a given right ascension $(\alpha)$ and
declination $(\delta)$ is specified with a model describing the intrinsic Stokes
$I(\nu), Q(\nu), U(\nu), V(\nu)$ flux densities of the source as a function of
frequency $\nu$. These flux densities are then converted to linearly polarized
coherencies at each channel frequency using Equation
\ref{VisModelLin}; 
this is also Equation 4.55 of \cite{TMS3}. 
These intrinsic receptor coherencies are then formed into the
$2\times2$ 
matrix  
\begin{equation}
    B(\nu) = 
    \begin{bmatrix}
        B_\mathrm{XX} & B_\mathrm{XY} \\
        B_\mathrm{YX} & B_\mathrm{YY}
    \end{bmatrix}
\end{equation}
which is computed for each simulated channel with frequency $\nu$; the flux
densities of simulated sources are assumed to be constant over time.

The antenna voltage responses of each linear polarization shown in
Figure~\ref{patt_volt} will scale the intrinsic coherencies $B(\nu)$ of a point
source differently for each timestamp and baseline. For an antenna $a$, the
value of the voltage response at the position of the source at
$(\alpha, \delta)$ with direction cosine coordinates $(l, m)$ relative to the
phase center is scaled and rotated to coordinates $(l_a', m_a')$. The
coordinates $(l_a', m_a')$ are derived from
\begin{equation}
    \begin{bmatrix}
        l_a'(t, \nu) \\
        m_a'(t, \nu)
    \end{bmatrix}
    =
    \left(\frac{\nu}{\nu_\mathrm{ref}}\right)
    \begin{bmatrix}
        \cos \psi_a(t) & \sin \psi_a(t) \\
        -\sin \psi_a(t) & \cos \psi_a(t) \\
    \end{bmatrix}
    \begin{bmatrix}
        l \\
        m
    \end{bmatrix},
\end{equation}
where the fraction $\frac{\nu}{\nu_\mathrm{ref}}$ radially scales the $(l, m)$
coordinates from the reference frequency $(\nu_\mathrm{ref})$ of the beam images
in Figure~\ref{patt_volt} (1510\,MHz) to the frequency of interest and the
source position is rotated (in an \textit{anti}-clockwise direction) by the
parallactic angle $\psi_a(t)$ of the antenna $a$ at time $t$. 

The complex-valued beam voltages $E_\mathrm{XX}(l_a', m_a')$,
$E_\mathrm{XY}(l_a', m_a')$, $E_\mathrm{YX}(l_a', m_a')$
and $E_\mathrm{YY}(l_a', m_a')$ for an antenna are obtained from a bi-cubic
interpolation of the beam voltage patterns in Figure~\ref{patt_volt}. These are
then formed into the $2\times2$ 
matrix 
\begin{equation}
    E_a(t, \nu, l, m) = 
    \begin{bmatrix}
        E_\mathrm{XX}(l_a', m_a') & E_\mathrm{XY}(l_a', m_a') \\
        E_\mathrm{YX}(l_a', m_a') & E_\mathrm{YY}(l_a', m_a')
    \end{bmatrix},
\end{equation}
for each simulated timestamp $(t)$, frequency $(\nu)$ and antenna
$(a)$. 
The coherency 
matrix  describing the beam-scaled flux density $F$ for a baseline between
two antennas $a_1$ and $a_2$ and a single point source $i$ is then
\begin{multline}
    F(t, \nu, l, m, a_1, a_2) = E_{a_1}(t, \nu, l, m) \times B(\nu) \times \\
                                E_{a_2}^\dagger(t, \nu, l, m)
                                \times e^{-2\pi j(u_{\nu}l_i +
                                    v_{\nu}m_i + w_{\nu}(n_i-1))}
\end{multline}
where the term $E_{a_2}^\dagger$ represents the complex conjugate and transpose
of the matrix $E_{a_2}$ 
and at a position given by direction cosines
$(l_i, m_i, n_i=\sqrt{1 - l_i^2 - m_i^2})$, on a baseline between antennas
$a_1$, $a_2$ with {\sc uv}-plane coordinates $(u_\nu, v_\nu, w_\nu)$ (measured
in wavelengths at the given time and frequency).
The flux density matrix $F$ describes the variation in
the measured coherencies caused by differences in per-antenna voltage patterns
in heterogeneous arrays (i.e. $F$ is a function of \textit{baseline}, and if the
antenna type of $a_1$ is MeerKAT and $a_2$ is SKA then $E_{a_1} \neq E_{a_2}$).
It further includes the effect of variations due to beam asymmetries (i.e. $F$
is a function of \textit{time}, and $E_\mathrm{XX}(l', m')$ etc. will vary as a
source position rotates with parallactic angle in a non-symmetric voltage
pattern).

When simulating $N$
point sources in the field the visibility for a given baseline, frequency
and timestamp is
\begin{equation}
V(t, \nu, a_1, a_2) = \sum_{i=1}^{N}{\frac{F_i(t, \nu, l_i, m_i, a_1, a_2)}{n_i}}.
\end{equation}

\section{Simulated Examples\label{Examples}}
Simulated noiseless data sets were imaged in Obit Task MFBeam which was given
the complex beam models described in Section \ref{Simulations} for the
MeerKAT and SKA antennas in all four Stokes correlations (XX,YY,XY,YX)
on a fine grid of frequencies. 
MFBeam uses faceting to correct for the ``w'' term and multiple
frequency bins are imaged independently but CLEANed jointly to accommodate
the spatially variant sky spectral distributions and antenna gain
patterns. 
The interpolated locations in the beam patterns were updated whenever
the parallactic angle changed by more than 0.25$^\circ$.
The ``perfect'' beam shape is a symmetrized version of that for
MeerKAT \citep{DEEP2}. 

Since the positions of sources are generally in a random location in a
facet and the dynamic range desired is high, the Obit autoCenter
feature is used in which sources with peaks in excess of a given
value, here 10 mJy/beam, are imaged centered in their own facets and
``unboxes'' are placed around the corresponding locations of other facets.
The depth of any major cycle is restricted to no deeper than 75\% of
the initial peak residual flux density.

For comparison, the data were also imaged in the same way by MFImage
in which beam corrections are not made; these are labeled
``Uncorrected'' in plots.
A ``Perfect'' noiseless, dataset with no beam corruptions applied to
the data was also imaged in MFImage to distinguish residual artifacts
from the beam corrections from those due to the basic imaging.
MFBeam and MFImage were run in an identical manner except for the
calculation of the instrumental response to the CLEAN components. 

Imaging was done on a Linux workstation running RHEL 7 with 16 x Intel(R)
Xeon(R) CPU E5-2687W~0 @ 3.10GHz cores and 256 GByte of RAM, of which
100 GByte was configured into a RAM disk which was used for output and
scratch files.
This machine supports AVX but does not have a GPU.
\subsection{Realistic Sky\label{SKADS}}
The Square Kilometer Array Design Study (SKADS) developed a suite of simulations
of the radio sky referred to collectively as the \textit{SKADS Simulated Skies}
(S$^3$). The S$^3$ Semi-Empirical extra galactic (S$^3$-SEX; \cite{wilman08})
simulation was constructed by drawing radio sources at random from observed and
extrapolated radio luminosity functions for various radio source populations and
positioning them with regard to a realistic model of their spatial clustering.
The S$^3$-SEX \textit{galaxies} simulation catalog contains positions in a
sky area of $20^\circ\times20^\circ$ and Stokes I flux densities derived from
a model of the frequency dependence for each galaxy population at 5
frequencies between 0.151 and 18\,GHz to a limit of 10\,nJy. 
A region containing 2401 galaxies within a $1^\circ$ radius to a
1400\,MHz flux density limit of $100\,\mu$Jy was used to provide a
realistic input sky model for generating noiseless visibilities with the
heterogeneous MeerKAT+ configuration. 
Each of the simulated point sources was fitted with a spectrum including a
spectral index and up to 2 curvature terms.

This realistic sky simulation was imaged with and without beam corrections.
Imaging consisted of a CLEAN of 100,000 components reaching a depth
of $\sim$10 $\mu$Jy/beam using a loop gain of 0.1.
The beam corrections resulted in the effective beam being the MeerKAT,
nominal, symmetric pattern \citep{DEEP2}.

\begin{figure*}
\centerline{ 
  \includegraphics[height=2.5in]{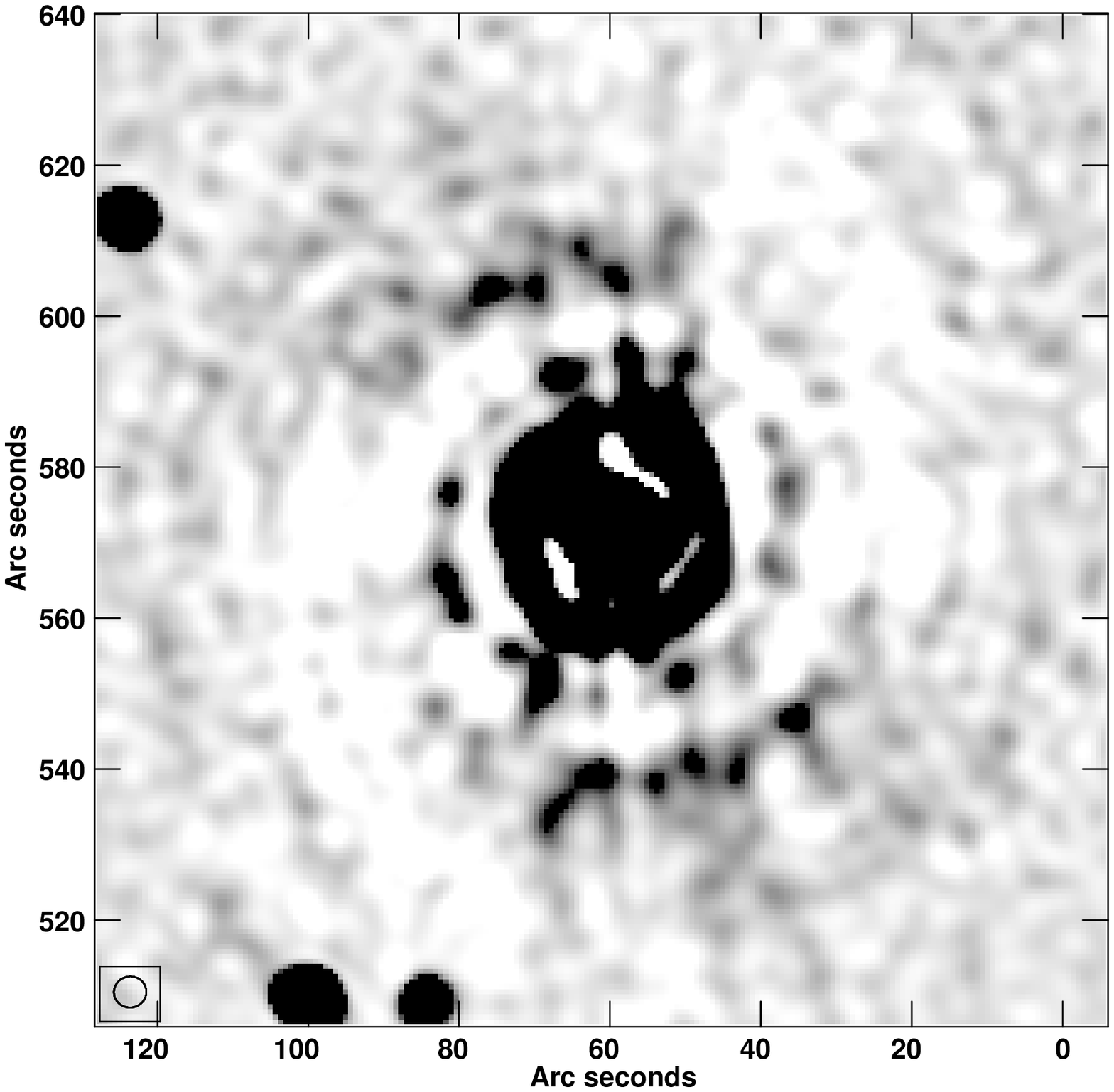}
  \includegraphics[height=2.5in]{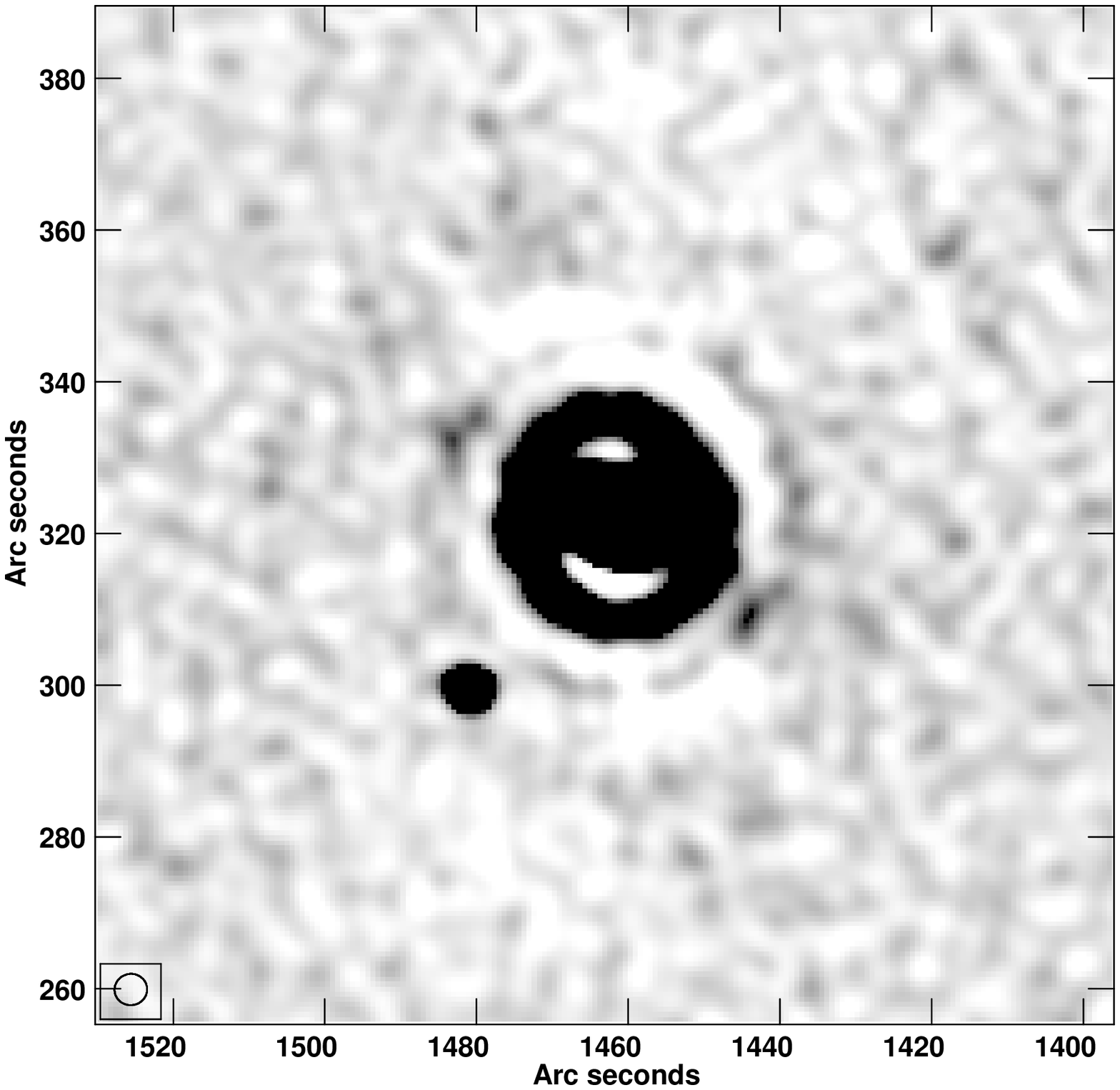}
  \includegraphics[height=2.5in]{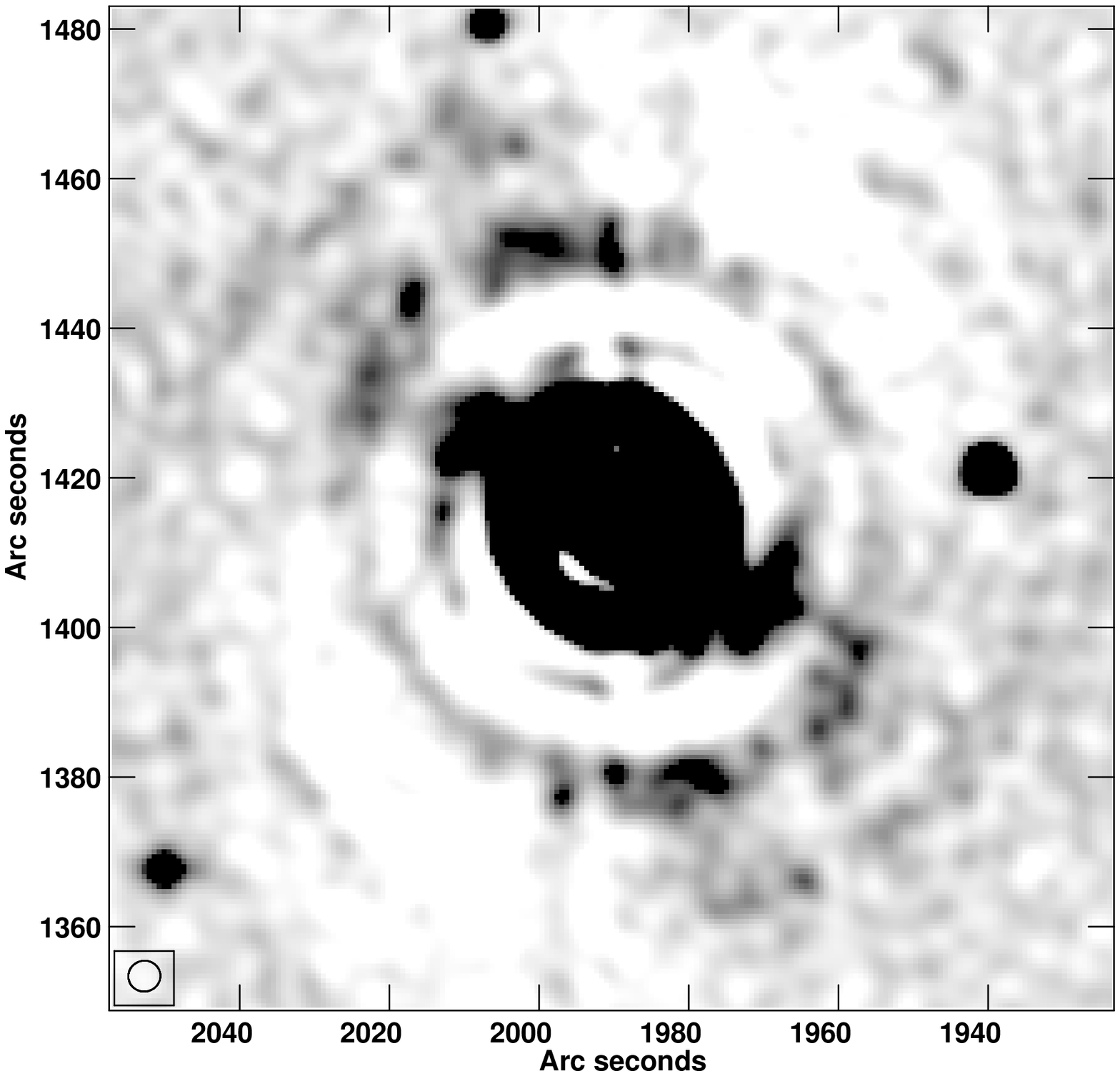}
}
\centerline{ 
  \includegraphics[height=2.5in]{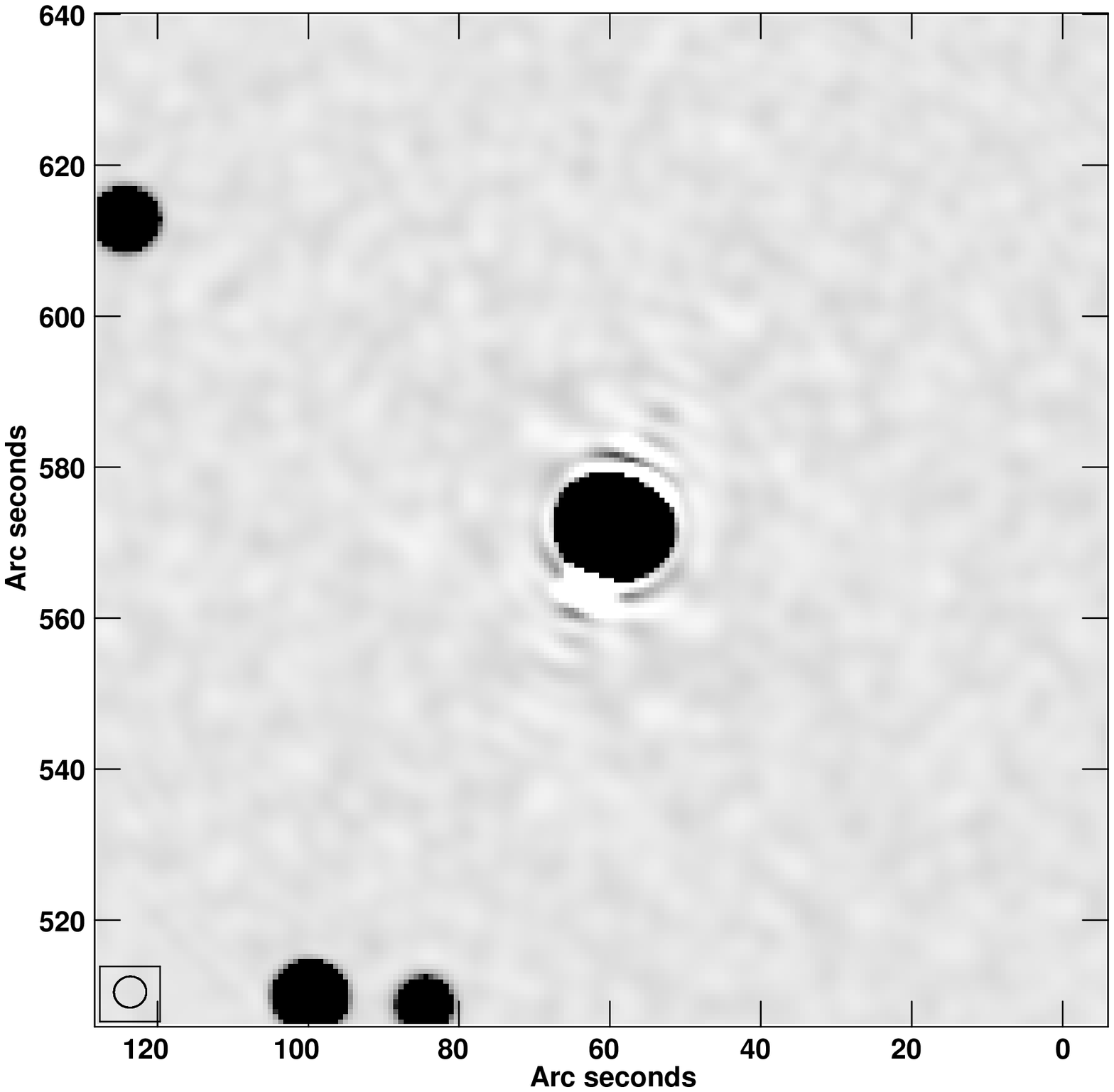}
  \includegraphics[height=2.5in]{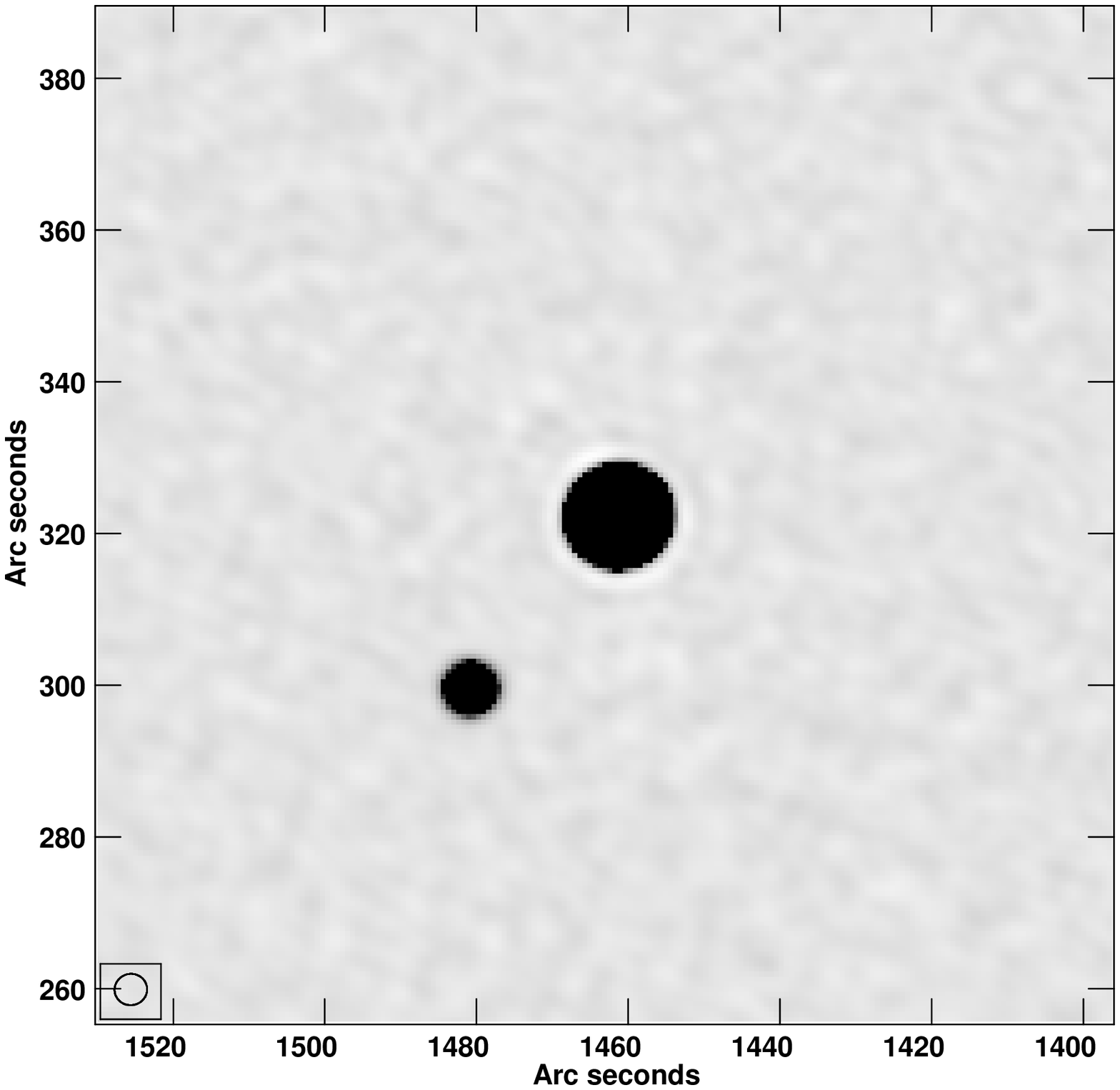}
  \includegraphics[height=2.5in]{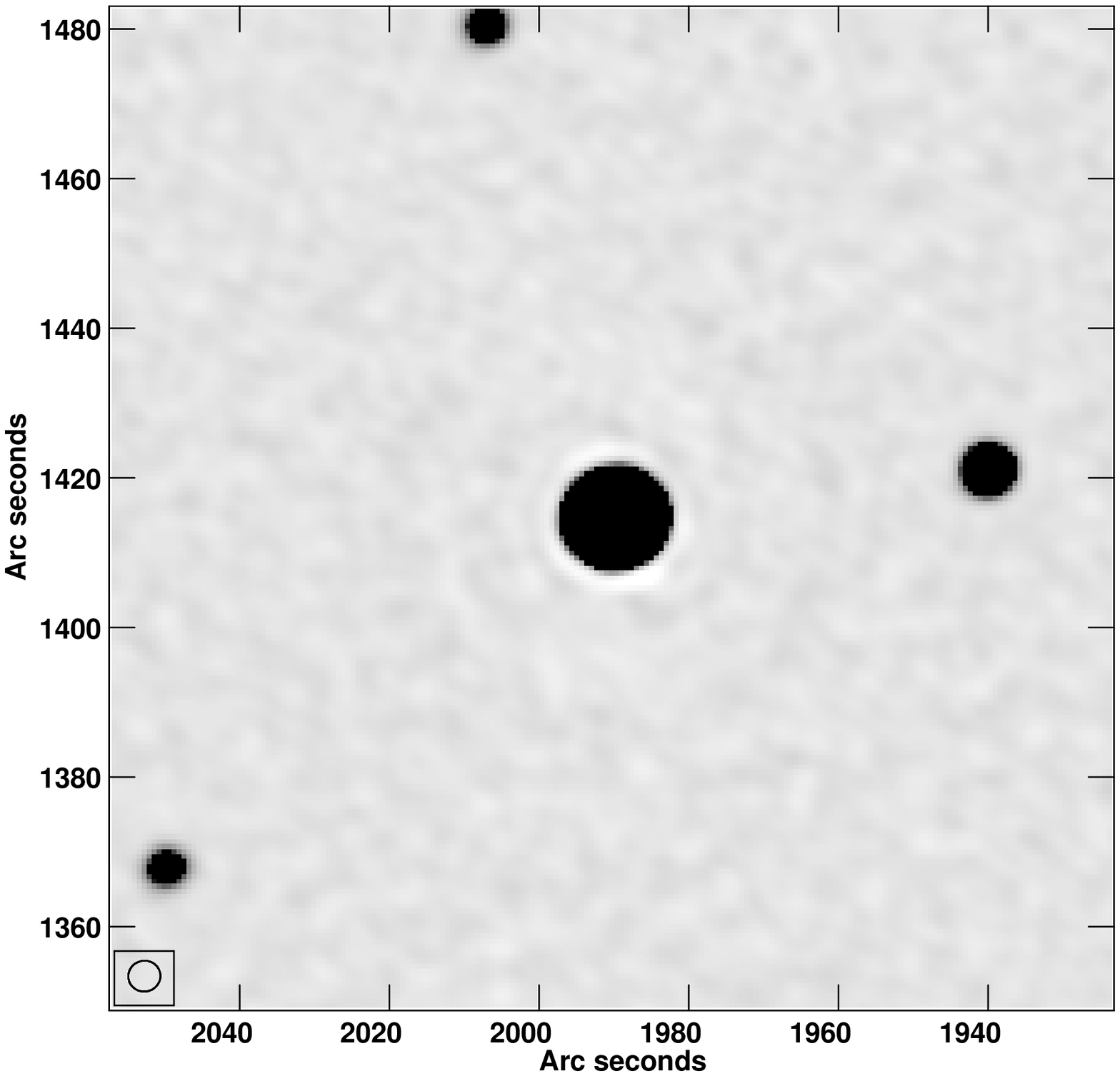}
}
\centerline{ 
  \includegraphics[height=2.5in]{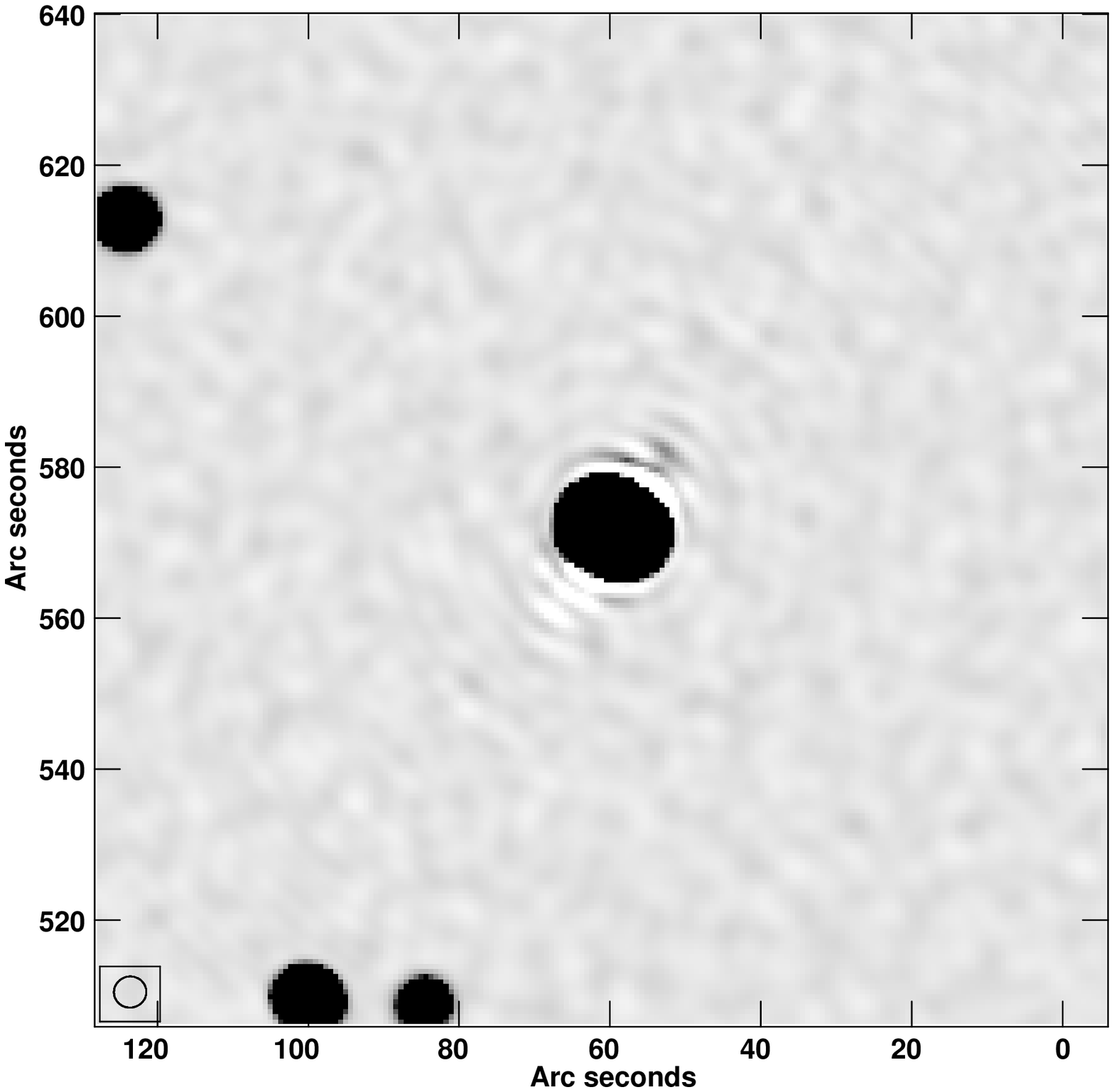}
  \includegraphics[height=2.5in]{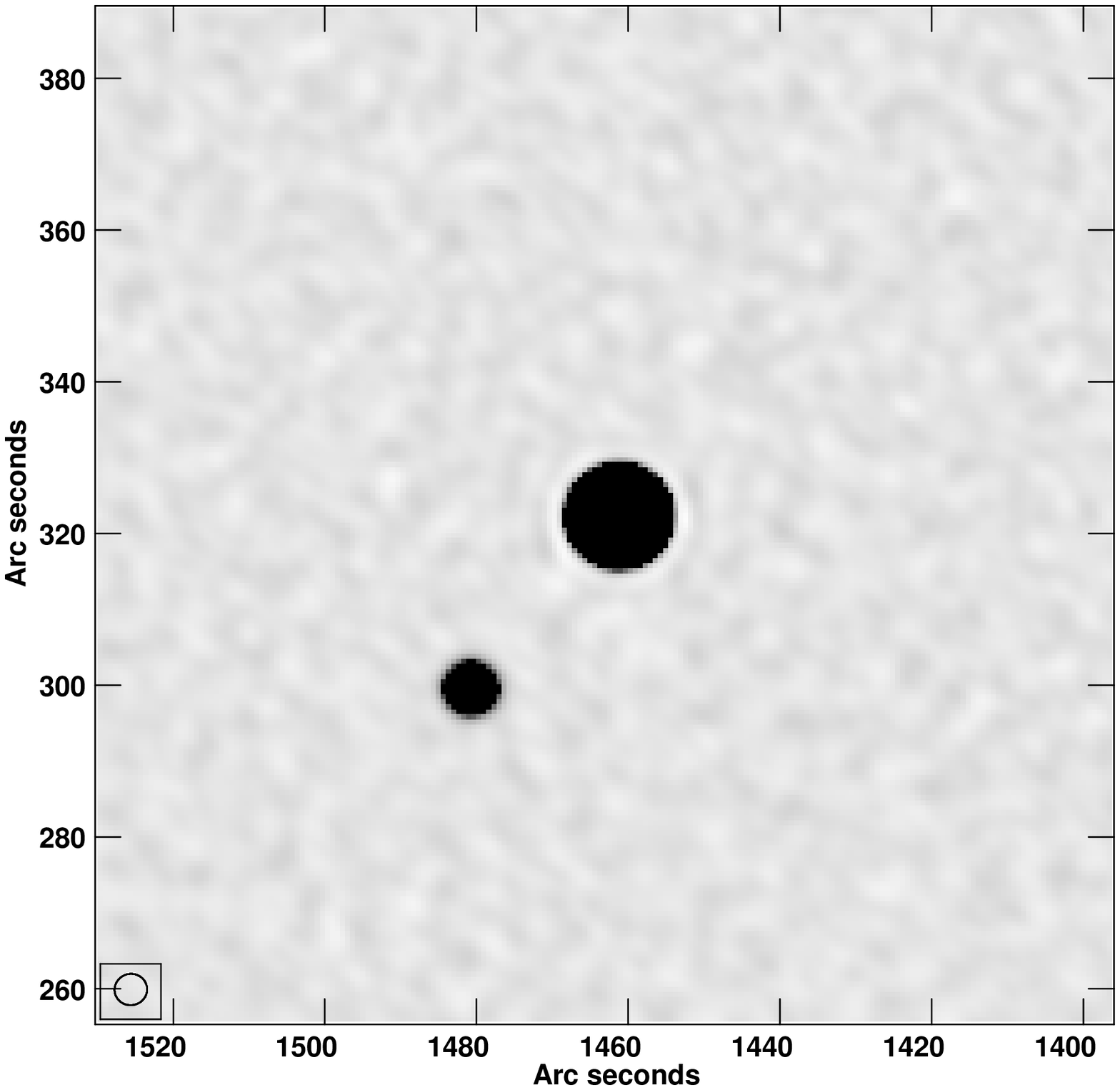}
  \includegraphics[height=2.5in]{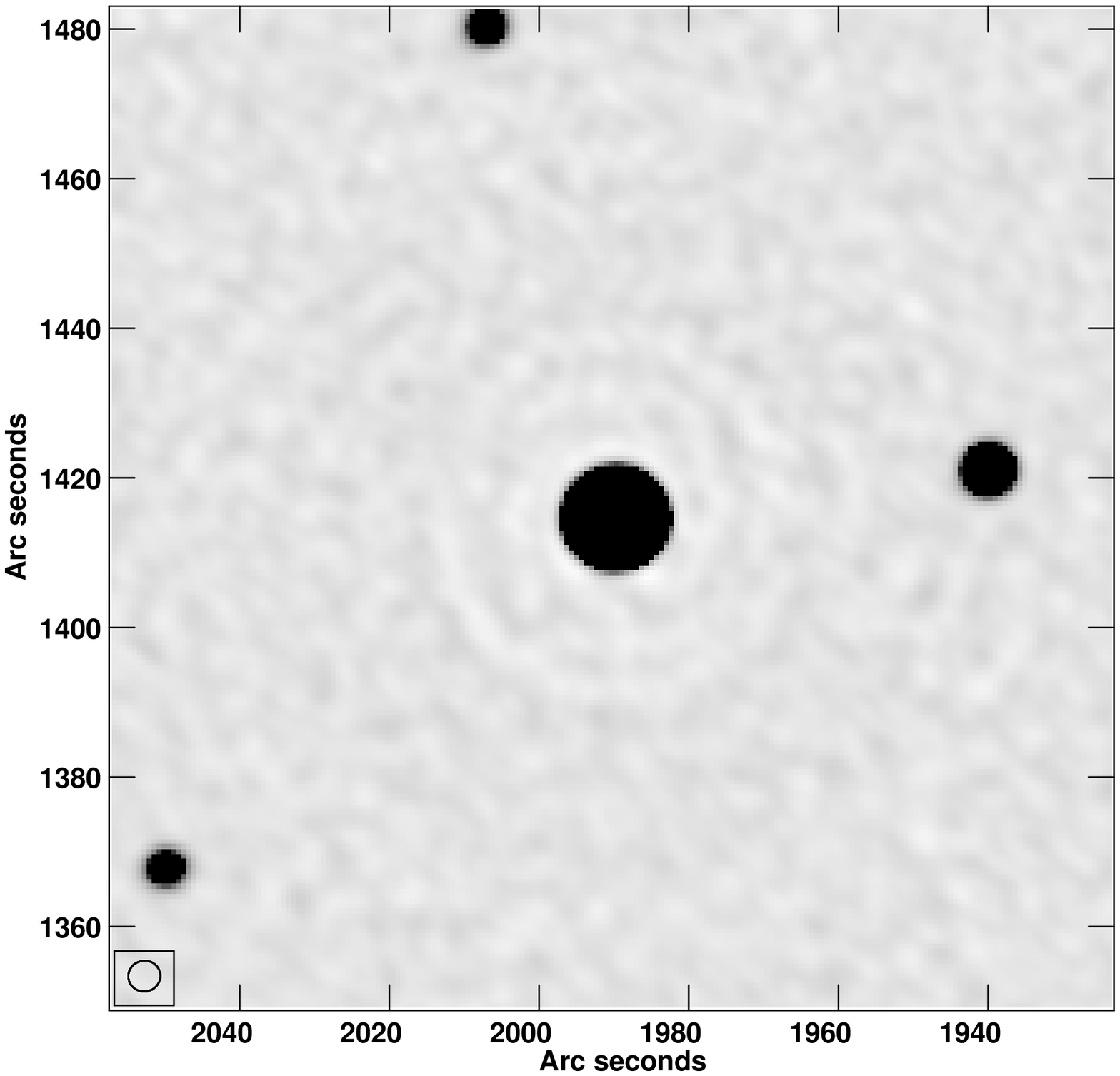}
}
\caption{Reverse gray scale of the Stokes I CLEAN restored image in
  regions around 3 brighter sources in the SKADS simulation.\\
{\bf Top Row:} Imaged without taking beam shapes into account. 
CLEANing was autoCentered with auto windowing.
The pixel range displayed is -5 $\mu$Jy to +15 $\mu$Jy.
Off--source RMS noise is 2.58 $\mu$Jy/bm.
The resolution is shown in the box in the lower left.\\
{\bf Middle row:} Like top row but taking beam shapes into account. 
Off--source RMS noise is 0.60 $\mu$Jy/bm.\\
{\bf Bottom row:} Like top row but with data simulated using a common, symmetric
  beam pattern, i.e. uncorrupted by beam shapes.
Off--source RMS noise is 0.79 $\mu$Jy/bm.
}
\label{SKADSImage}
\end{figure*}

Without beam corrections, the off--source RMS was 2.58 $\mu$Jy/bm and with
beam corrections 0.60 $\mu$Jy/bm.
The peak in the image is 0.55 Jy so the nominal dynamic ranges over
the whole field of view were 0.92$\times 10^6$ with beam correction
and 0.21$\times 10^6$ without. 
Beam correction made further dynamic range improvements in the regions
around bright sources.
Cutouts around some of the stronger sources imaged by MFImage (no beam
correction) are shown in Figure \ref{SKADSImage} {Top Row} and
the same sources imaged by MFBeam (beam corrected) in Figure
\ref{SKADSImage} {Middle Row}
with the same stretch. 
The beam corrections greatly reduced the near in artifacts.

This simulation was also repeated except using the same, symmetric beam shape
for all antennas which should not introduce artifacts.
Imaging of this data with MFImage should show only the residual
artifacts from imaging. 
Such results are shown in Figure \ref{SKADSImage} {Bottom Row}
and should be compared with Figure \ref{SKADSImage} {Middle Row}.
The bulk of the residual artifacts in Figure \ref{SKADSImage} {
Middle Row} appear not to be the result of uncorrected beam effects.
The RMS in the image from data without beam corruptions was 0.79
$\mu$Jy beam$^{-1}$.  

\begin{figure}
\centerline{ 
  \includegraphics[height=3.5in,angle=-90]{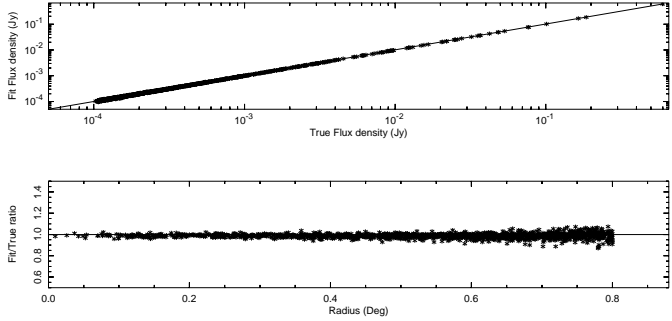}
}
\caption{Comparison of recovered ('Fit') source flux densities to true
  values for the SKADS simulation.
Top panel, the log ``fitted'' v. the log true values.
The line shows the 1:1 values.
The bottom panel gives the ratio of fit to true flux densities as a
function of the distance from the pointing center.
}
\label{SKADSFlux}
\end{figure}

A comparison of the derived flux densities (Obit/FndSou) after a
correction for the primary beam attenuation to the initial model
values is shown in Figure \ref{SKADSFlux}. 
This figure shows that the flux densities in the input model are well
recovered.

\begin{figure}
\centerline{ 
  \includegraphics[height=3.5in,angle=-90]{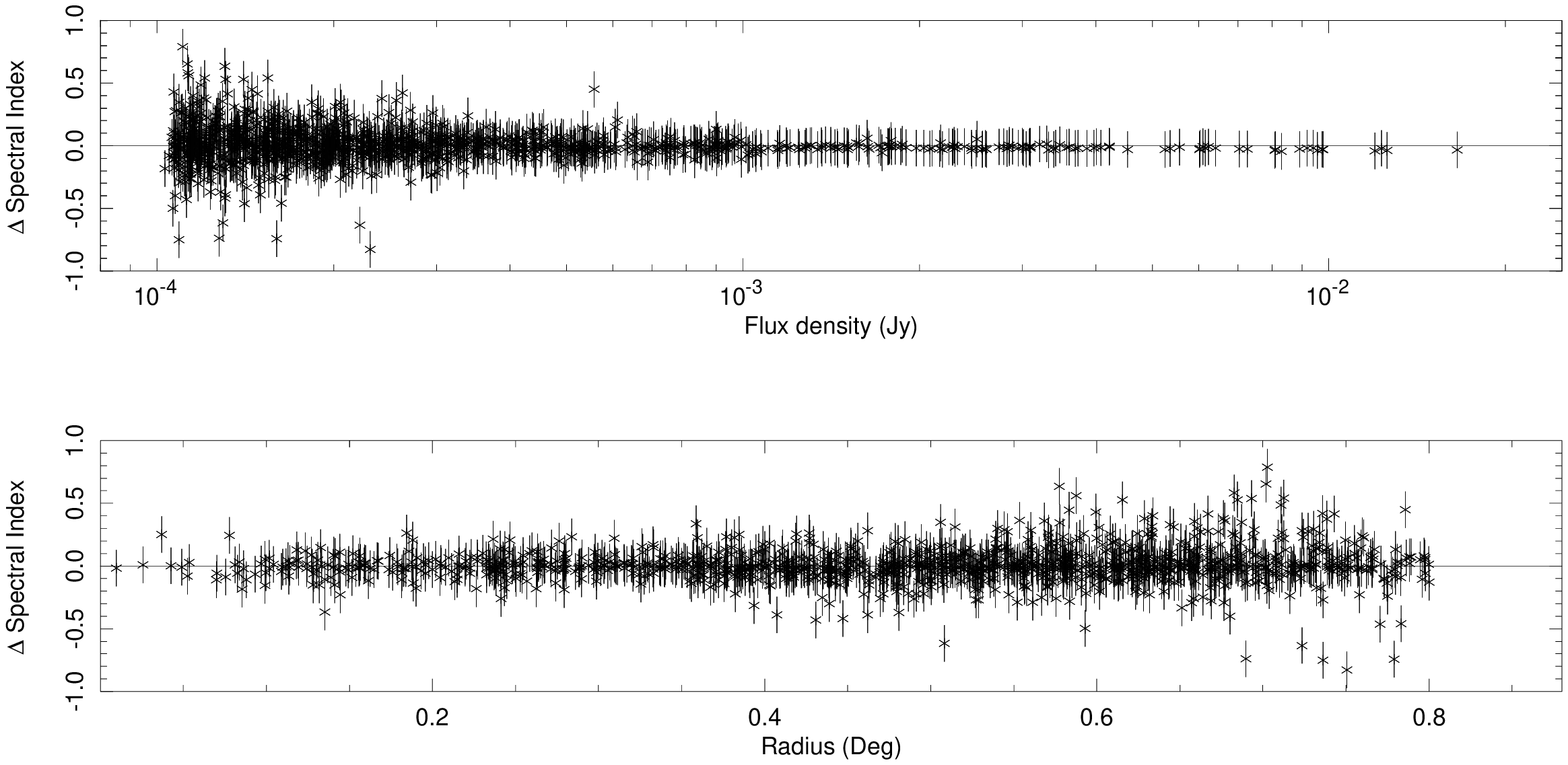}
}
\caption{Difference between the fitted and true spectral index as a
  function of flux density and distance from the pointing center with
  fitting error bars for the SKADS simulation.
Top panel, the fitted spectral index difference v. log flux density.
The line shows the expected value (0.0).
The bottom panel gives the spectral index difference as a function of
the distance from the pointing center. 
}
\label{SKADS_SI}
\end{figure}

The input model consisted of point sources with a variety of spectra.
The derived image had spectra fitted in each pixel with flux density
at the reference frequency, spectral index and up to 2 curvature
terms. 
The fitted spectral indices after correction for the primary beam gain
are compared with the input model values in Figure \ref{SKADS_SI}.
The bulk of the fitted values are close to the input model values but
with some scatter which increases for weaker sources and
increasing distance from the pointing center.


Timing tests of imaging with corrections (MFBeam), without corrections
(MFImage) and without beam corruptions (MFImageSym) were run as above
but with a minimum flux density of 100 $\mu$Jy beam$^{-1}$.
These tests were identical except for the interferometer model
calculation and the input data and are summarized in Table
\ref{timeTab}.

CLEAN deconvolution is a nonlinear process and even slight differences
can result in different paths to the solution.
This is obvious from Table \ref{timeTab} in the varying amounts of work
performed (columns ``Comp.'', ``Major'' and ``Facet'').
The difference in the run time between with and without correction is
a function of 1) differences in the amount of work done, 2)
differences in the cost of the interferometer model calculation and 3) the
fraction of the total time used by the interferometer model
calculation so the results in Table \ref{timeTab} are only suggestive.
In this case the run time for applying the corrections is increased by
50\% over not applying the corrections.
The larger difference in the CPU times shows this to be quite
sensitive to the number of CPU cores used, here 16.

Further note that this test did not use GPU enhancement for the
interferometer model calculation.
Use of the GPU model calculation available for MFImage shows that it is so
much faster than the CPU version that it frequently becomes a
relatively insignificant fraction of the total run time \citep{ObitMemo35}.
Similar gains are expected for a GPU version of the calculation used
in MFBeam.

\begin{table}
\caption{SKADS Timings}
\vskip 0.1in
\begin{center}
\begin{tabular}{lrrccc}   \hline
\hline
Test& clock &CPU&Comp.& Major&Facet\\
    &  hr   & hr &  &  &  \\
\hline
MFBeam    & 2.96 & 23.19 & 11,347 & 88 & 11,955 \\
MFImage   & 2.00 &  9.97 & 13,562 & 77 & 10,630 \\
MFImageSym& 2.24 & 11.23 & 11,806 & 81 & 12,743 \\
\hline
\end{tabular}
\end{center}
Notes: ``clock'' = wall clock time, ``CPU'' = CPU time, ``Comp.'' is
the number of CLEAN components used, ``Major'' is the number of major
CLEAN cycles and ``Facet'' is the number of facet imagings.
\label{timeTab}
\end{table}

%

\subsection{Unpolarized Grid}
A second simulation used a sample of nine unpolarized sources on a
grid covering the field of view to explore the spurious polarized
response introduced by the beam shapes.
This test consists of 2 parts:
\begin{itemize}
\item {\bf Uncorrected} \\
These data used separate, asymmetric, MeerKAT and SKA beams and are
corrupted by the beam effects.
These corruptions were ignored by imaging in MFImage.
\item {\bf Corrected} \\
The same corrupted data were also imaged in MFBeam with corrections in the
Stokes I.
The resultant dataset with the corrections for the instrumental
response were then imaged in Stokes Q, U and V without
further correction.
\end{itemize}

\begin{figure}
\centerline{ 
  \includegraphics[width=3.0in,angle=-90]{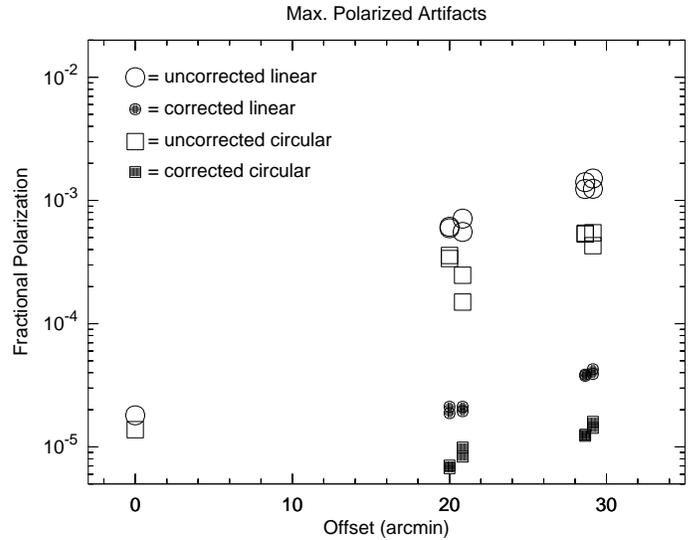}
}
\caption{Maximum fractional polarization artifacts as a function of
  distance from the pointing center 
derived from Table \ref{GridTab}. 
Open circles and squares are the uncorrected linear and circular
polarizations and the filled symbols are those after beam
corrections. 
}
\label{MaxArt}
\end{figure}

The ratios of the maximum linear and circular polarization artifacts
in the vicinity of each source to the total intensity flux density
shown in Table \ref{GridTab} are plotted in Figure \ref{MaxArt}.
As expected, the maximum fractional polarization artifacts are very
low near the pointing center and increase with increasing distance.
The corrections applied in this test reduced the maximum linearly
polarized artifacts by approximately a factor of 30 and somewhat more
for circular polarization.
Our speculation is that since Stokes V is the difference in small
values with linear feeds and Q and U are combinations of larger
values, residual errors in the corrections will have a smaller effect
on Stokes V.

\begin{table*}
\caption{Unpolarized Grid Test Artifacts}
\vskip 0.1in
\begin{center}
\begin{tabular}{lrrcccccc}   \hline
\hline
Source& dist &max I& Q$_{uncorr}$& Q$_{corr}$& U$_{uncorr}$& U$_{corr}$&  V$_{uncorr}$& V$_{corr}$ \\
     &' &mJy bm$^{-1}$& $\mu$Jy bm$^{-1}$& $\mu$Jy bm$^{-1}$& $\mu$Jy bm$^{-1}$& $\mu$Jy bm$^{-1}$& $\mu$Jy bm$^{-1}$& $\mu$Jy bm$^{-1}$ \\
\hline
 A &44 &56.3 & 13.9 -11.1 & 2.2 -0.4  & ~9.1 -70.0 & 0.4 -0.3 & ~7.1 -24.2 & 0.9 -0.1 \\
 B &30 &70.9 & ~9.0 -42.1 & 0.3 -0.4  & 21.1 -14.6 & 0.3 -1.5 & ~6.8 -24.0 & 0.5 -0.3 \\
 C &44 &56.2 & 29.3 -21.3 & 0.4 -2.4  & 84.7 -15.6 & 0.6 -0.3 & 16.8 -30.7 & 0.5 -0.8 \\
 D &31 &70.9 & 39.4 ~-5.1 & 0.2 -0.2  & 14.1 -10.1 & 1.5 -0.8 & ~7.0 -10.6 & 0.7 -0.1 \\
 E &0 &94.3 & ~1.0 ~-1.2 & 0.0 -0.0  & ~1.7 ~-1.3 & 0.0 -0.0 & ~0.5 ~-1.3 & 0.0 -0.0 \\
 F &31 &72.0 & 51.2 -10.9 & 0.3 -0.2  & 14.4 -11.2 & 1.4 -0.3 & 17.9 -12.2 & 0.4 -0.6 \\
 G &43 &55.6 & 16.5 -22.1 & 0.3 -2.1  & 68.8 ~-8.7 & 0.4 -0.6 & 30.1 -10.2 & 0.7 -0.2 \\
 H &30 &74.6 & ~8.6 -45.6 & 0.4 -0.2  & 20.8 -22.4 & 0.3 -1.4 & 26.6 ~-7.7 & 0.5 -0.3 \\
 I &43 &57.0 & 17.4 -18.8 & 2.2 -0.4  & 13.6 -80.2 & 0.5 -0.4 & 30.4 -11.8 & 0.7 -0.8 \\
\hline
\end{tabular}
\end{center}
Notes: ``dist'' is the distance from the pointing center, ``Max I'' is
the peak Stokes I and the following columns are Stokes Q, U and V
uncorrected and corrected for beam effects; 
each column contains the maximum and minimum within 1.1' in RA
  and dec. of each source.
\label{GridTab}
\end{table*}

\subsection{Polarized Grid}
A third simulation is similar to the second except that the grid of
sources were each partially polarized.
This test consists of 3 parts:
\begin{itemize}
\item {\bf ``Perfect'' data} \\
The simulations used the same, symmetric beam patterns for all
antennas. 
These should have no corruptions and were imaged by MFImage.
Residual artifacts are the results of the imaging rather than
limitations of the beam corrections.
\item {\bf Uncorrected} \\
These data used separate, asymmetric, MeerKAT and SKA beams and are
corrupted by the beam effects.
These corruptions were ignored by imaging in MFImage.
\item {\bf Corrected} \\
The corrupted data were also imaged in MFBeam with corrections in the
order, Stokes I, Q, U and V.
\end{itemize}

\begin{figure*}
\centerline{ 
  \includegraphics[height=2.2in]{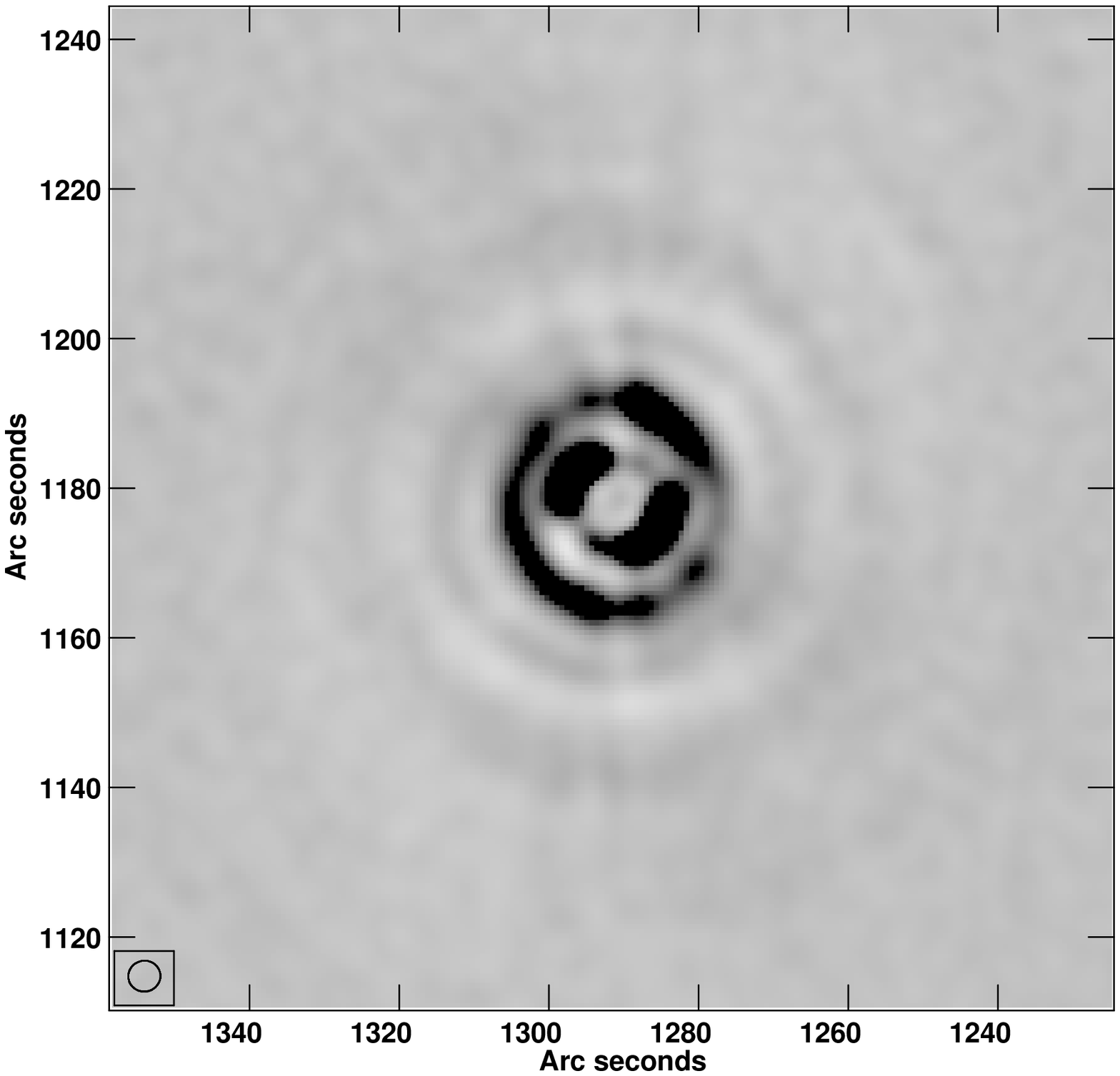}
  \includegraphics[height=2.2in]{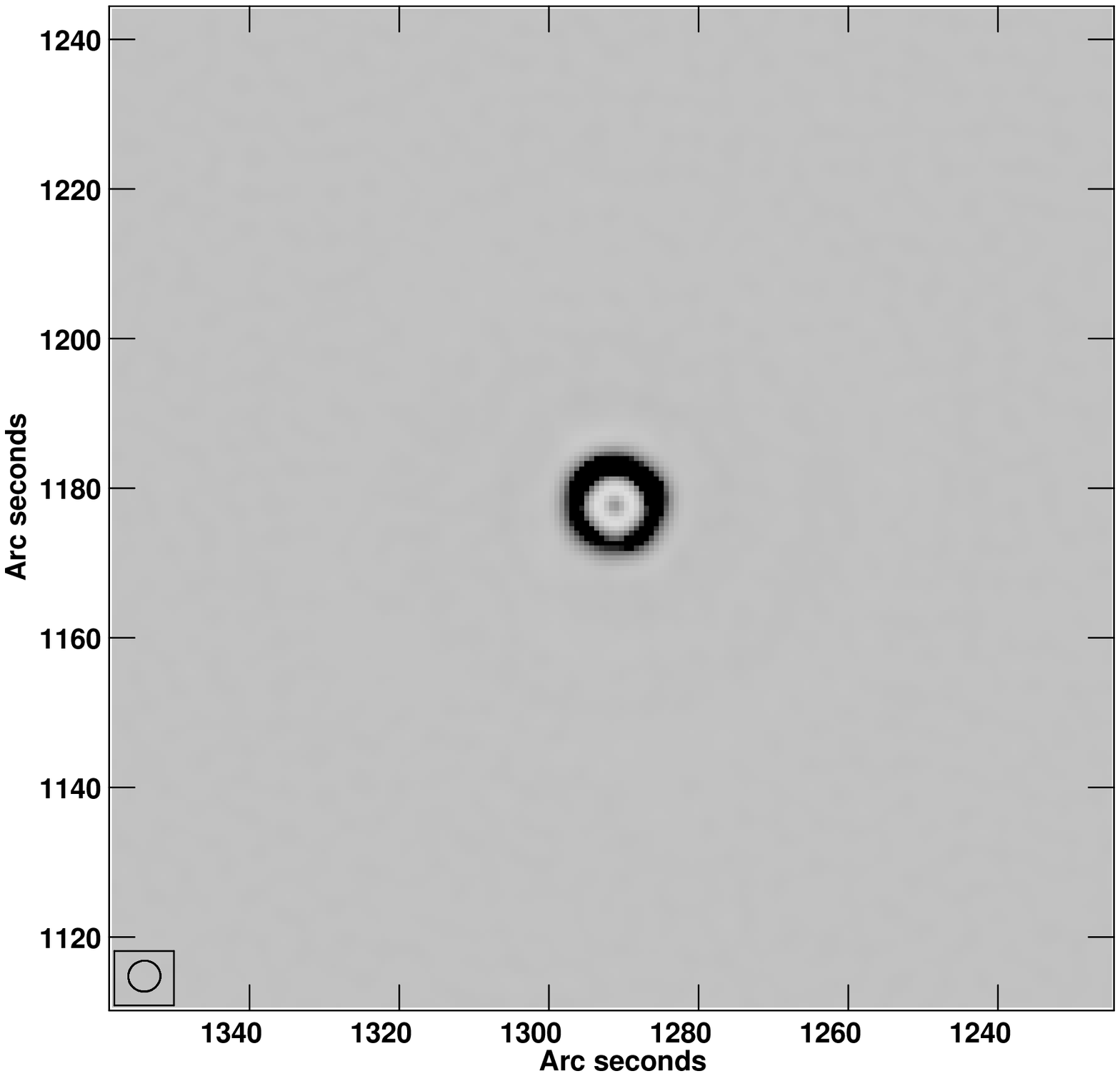}
  \includegraphics[height=2.2in]{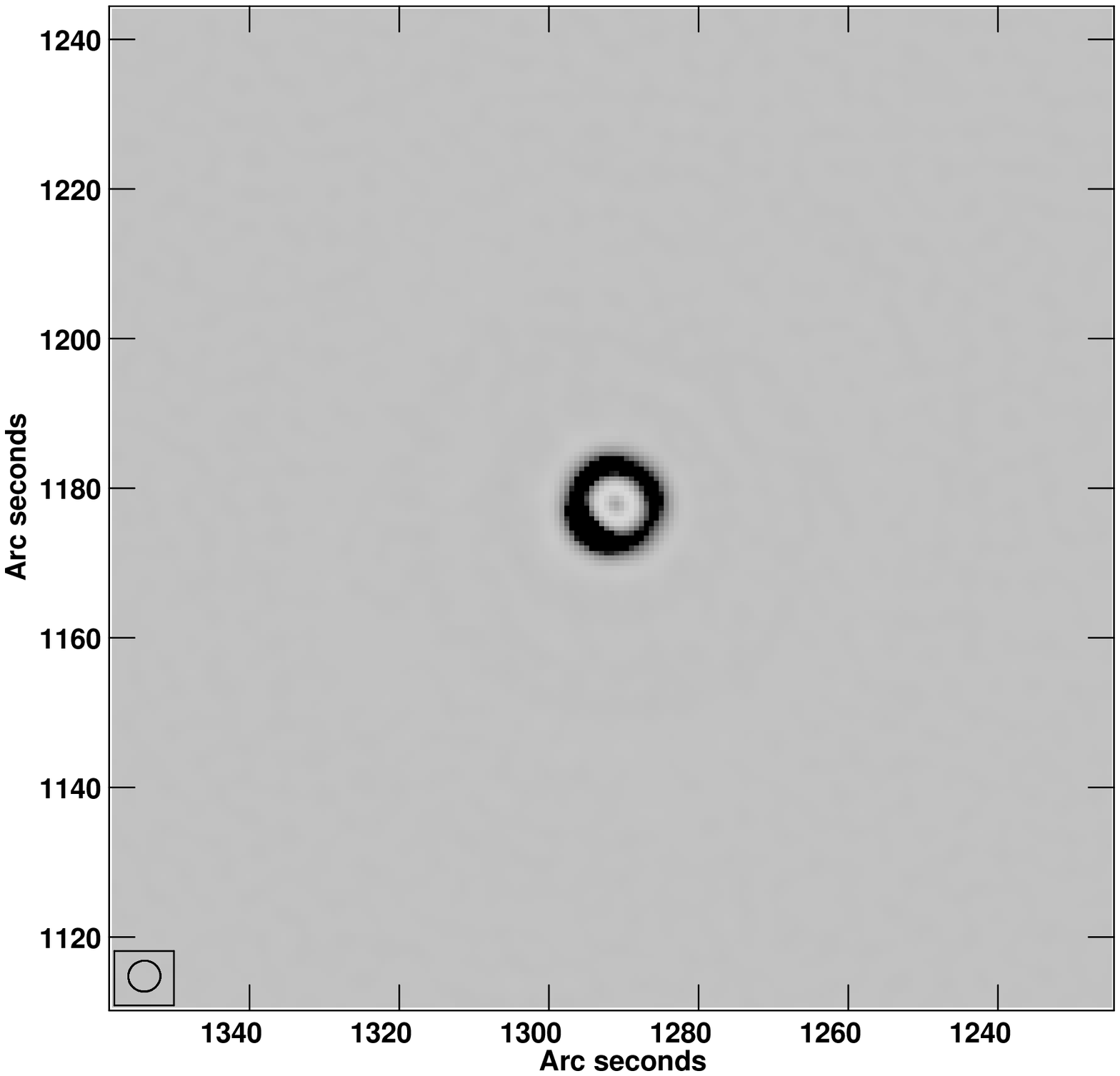}
}
\centerline{ 
  \includegraphics[height=2.2in]{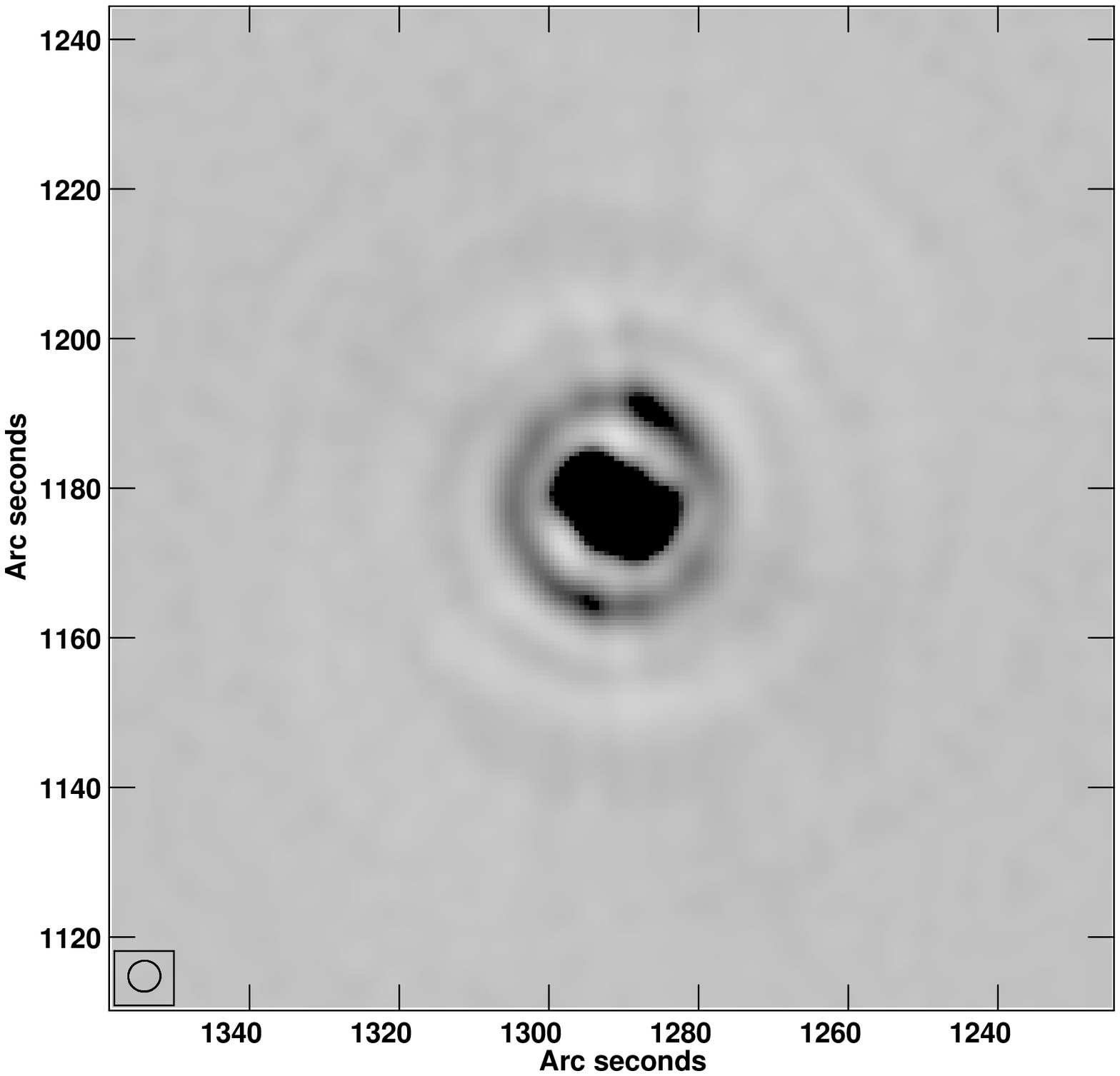}
  \includegraphics[height=2.2in]{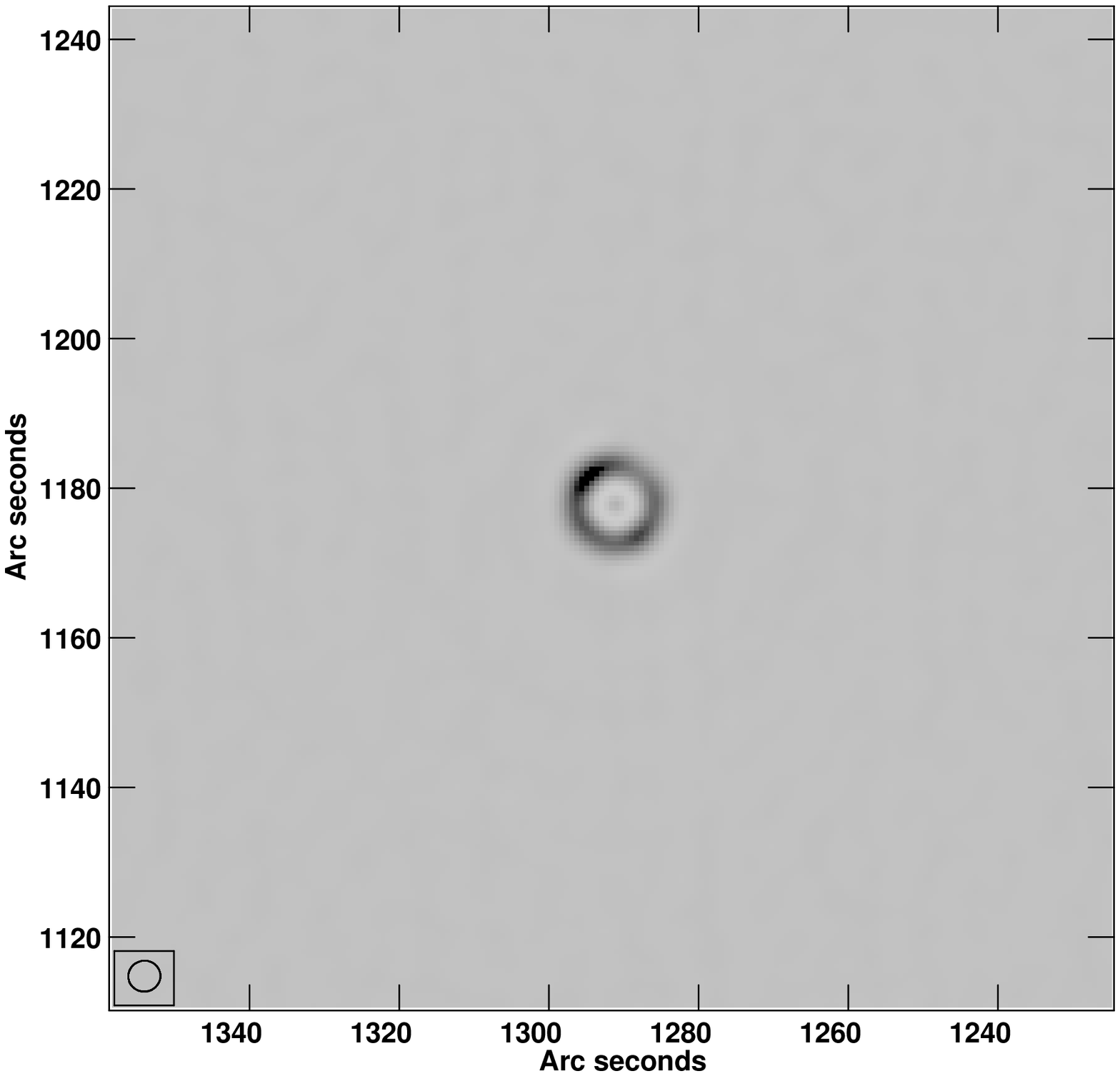}
  \includegraphics[height=2.2in]{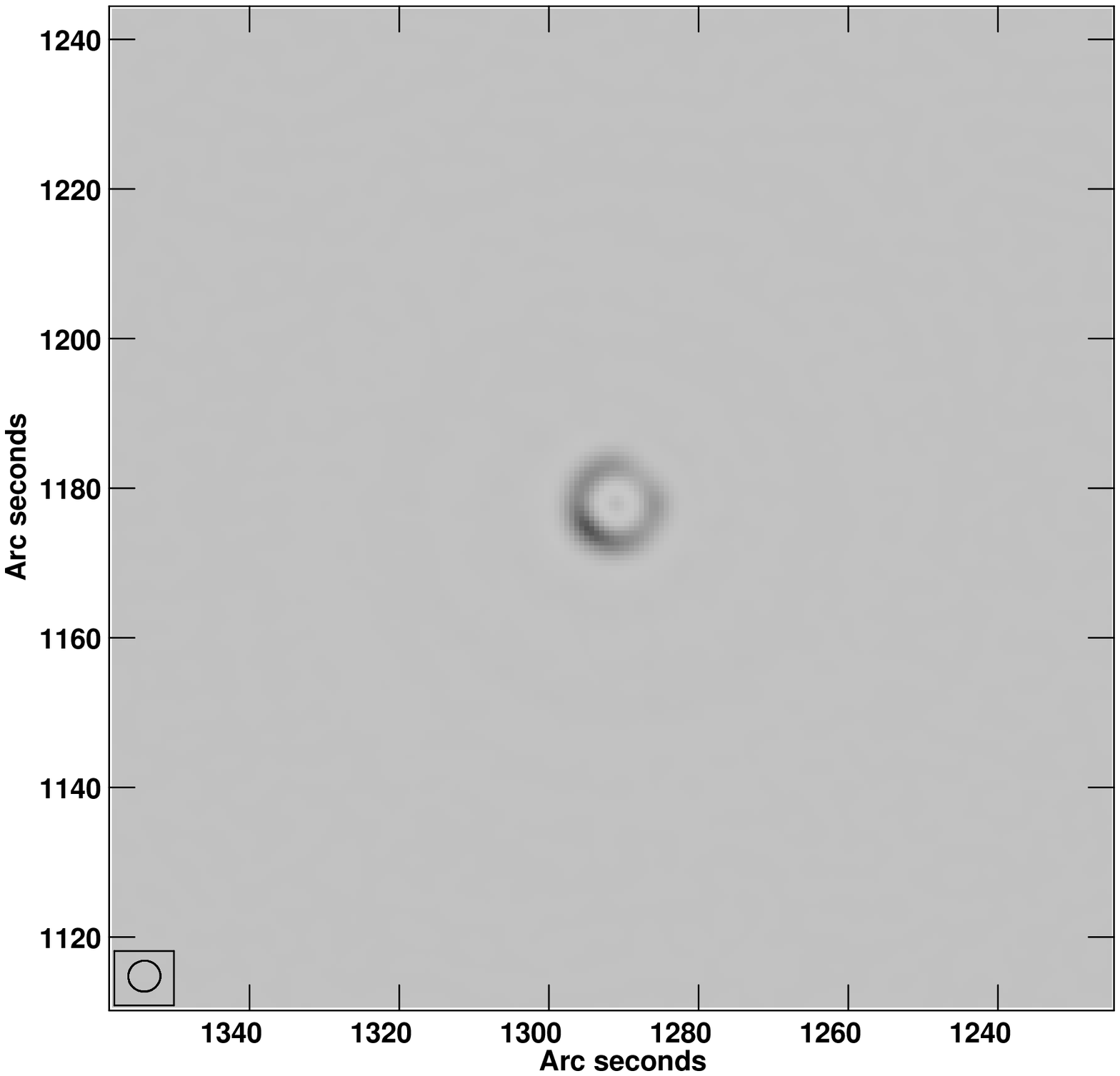}
}
\centerline{ 
  \includegraphics[height=2.2in]{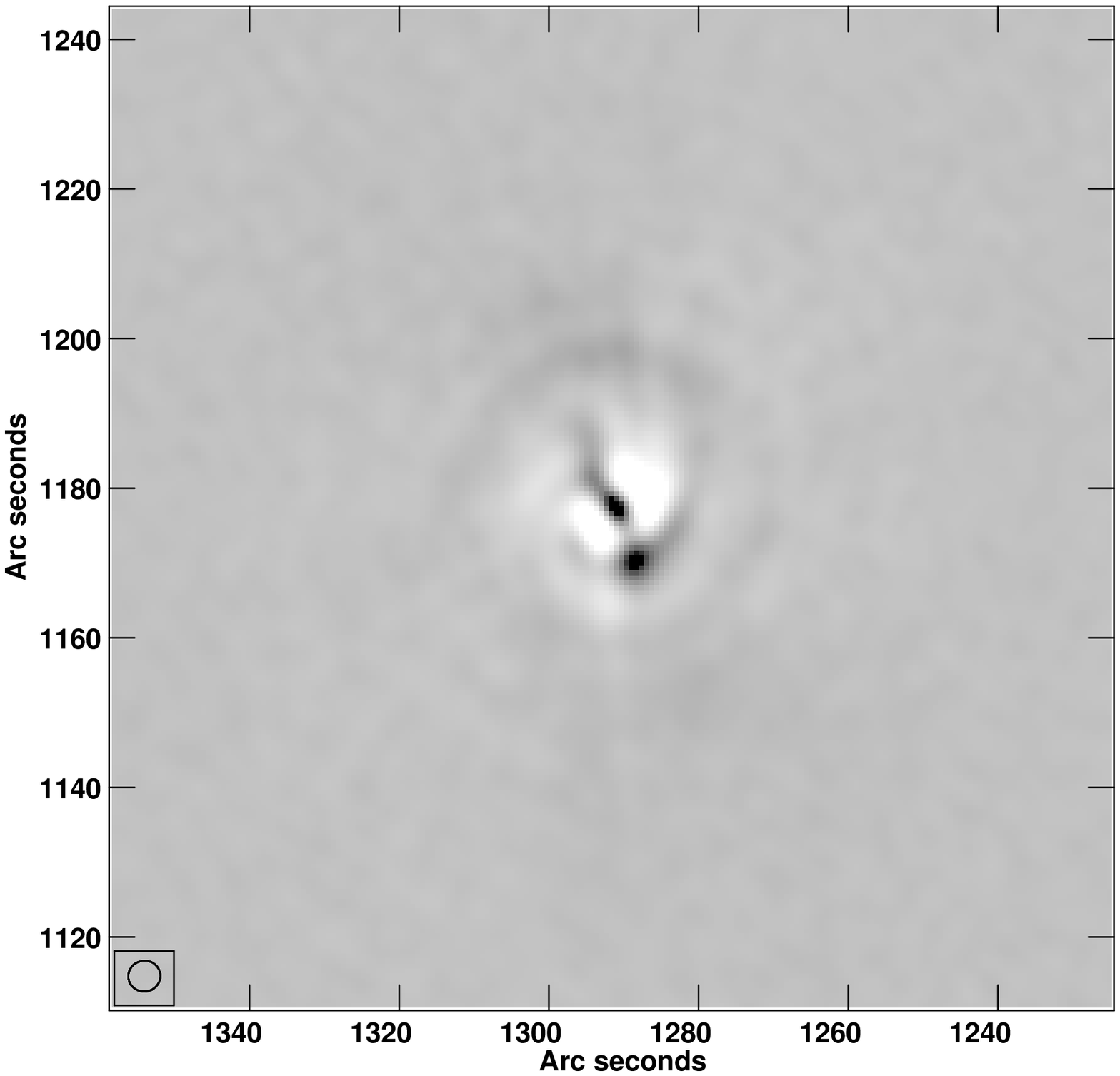}
  \includegraphics[height=2.2in]{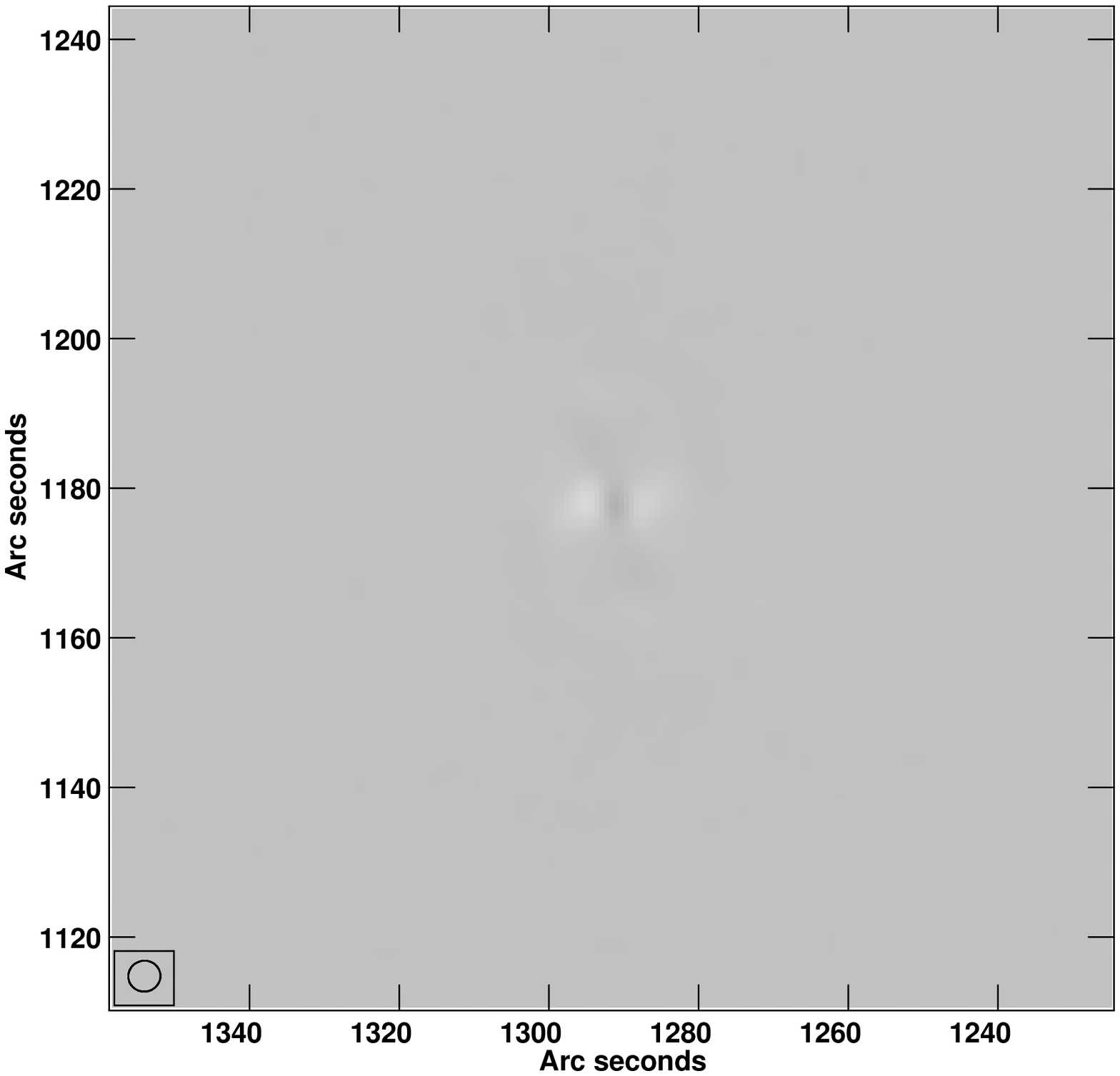}
  \includegraphics[height=2.2in]{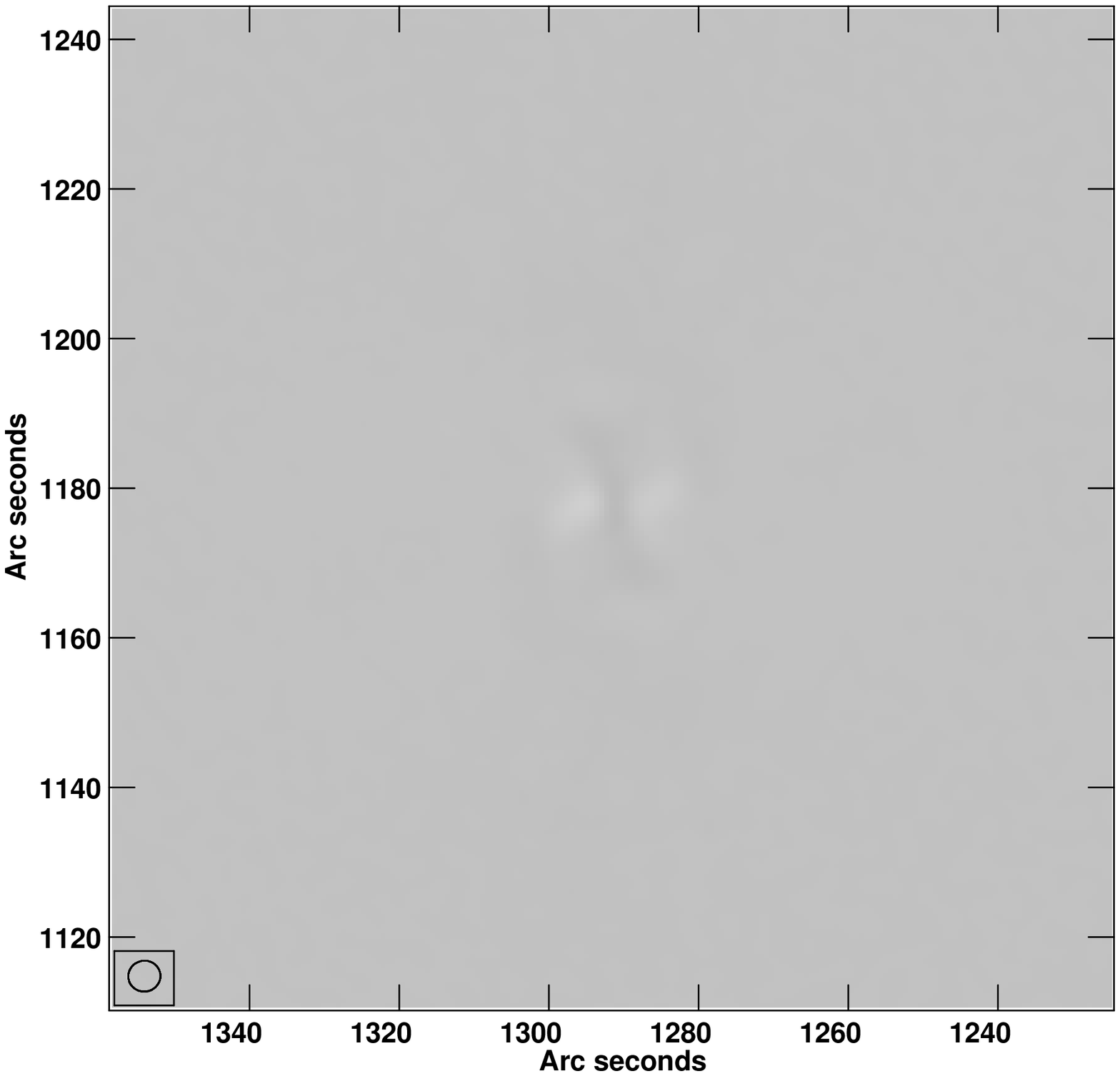}
}
\centerline{ 
  \includegraphics[height=2.2in]{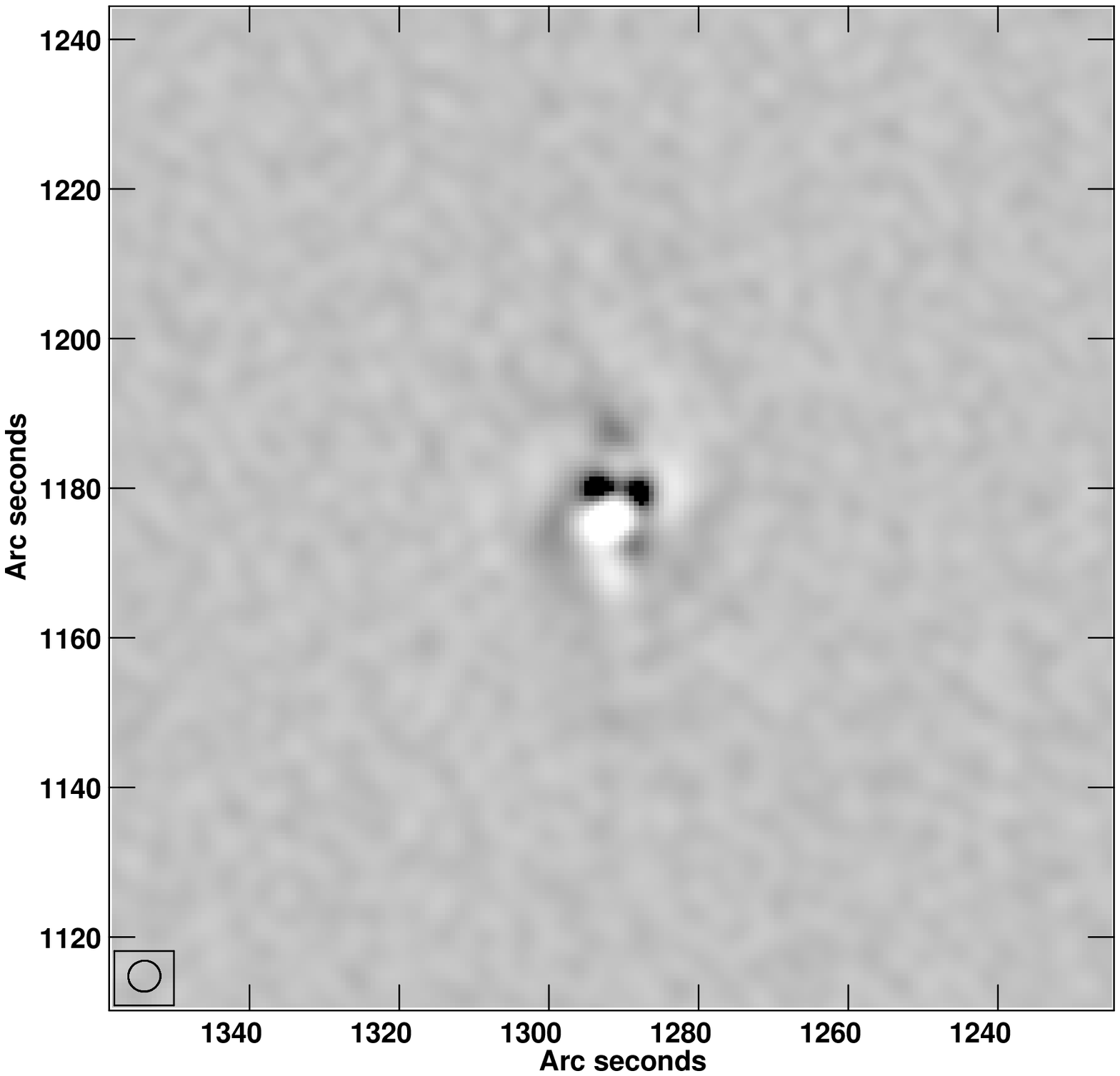}
  \includegraphics[height=2.2in]{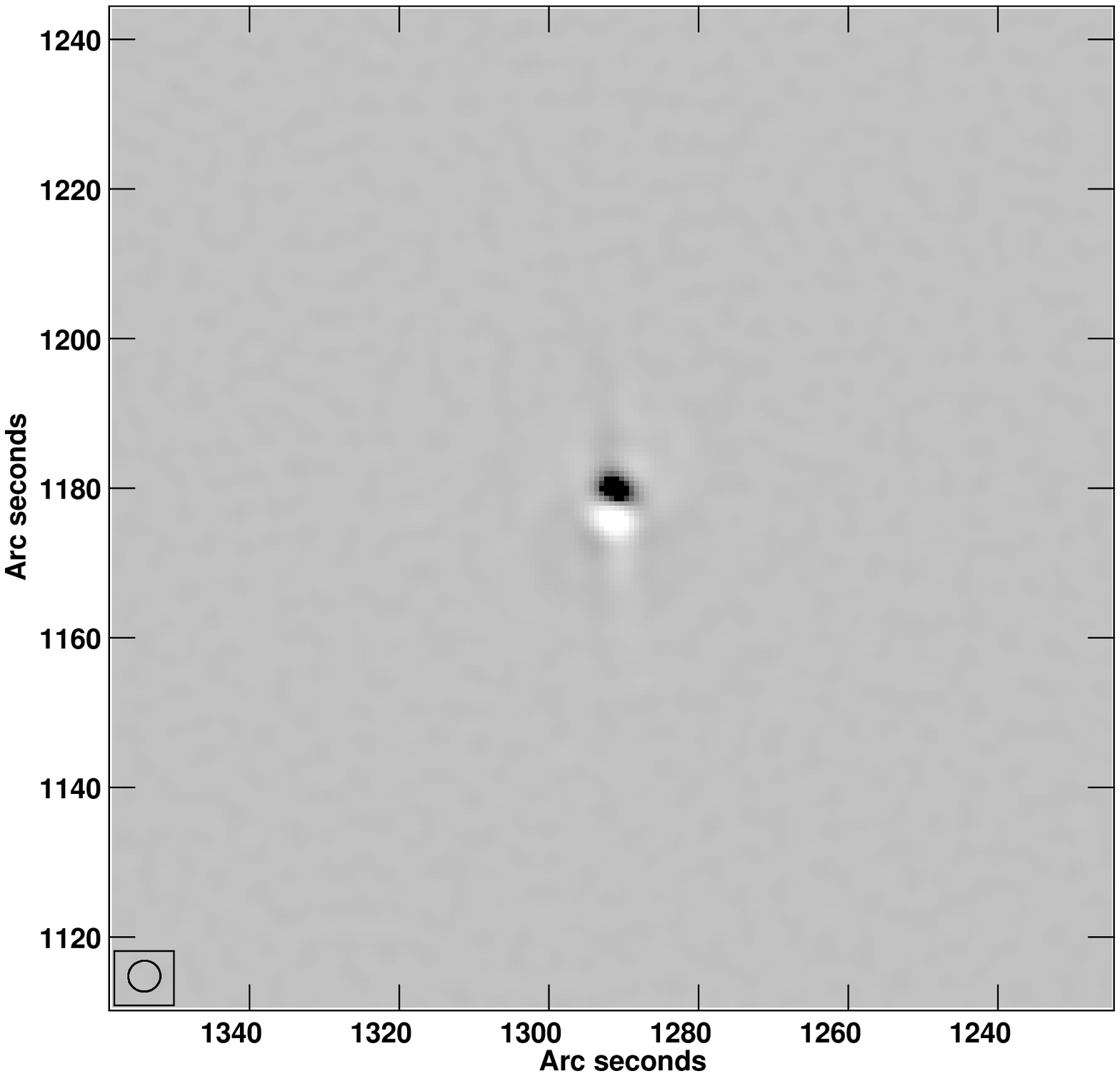}
  \includegraphics[height=2.2in]{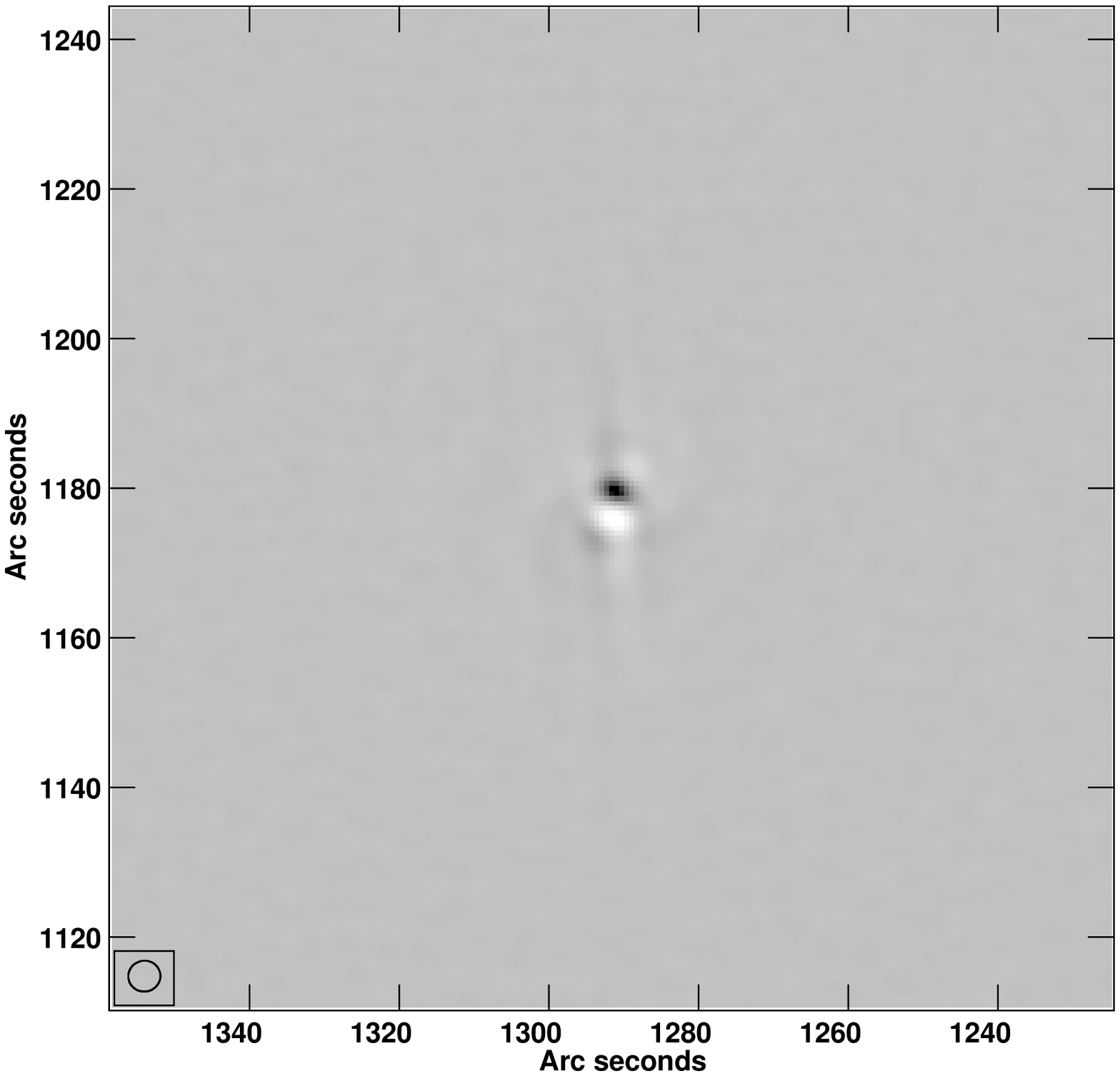}
}
\caption{Reverse gray scale of the residuals around source A.
Left column is the uncorrected image, center column is with
corrections and the right column is the ``Perfect'' image.
Top row is Stokes I, second row is Stokes Q, third row Stokes U and
bottom row is Stokes V.
The Stokes I pixel range displayed is $\pm\ 50\ \mu$Jy,Stokes Q \& U,
$\pm\ 25\ \mu$Jy, and for Stokes V $\pm\ 10\ \mu$Jy.
The resolution is shown in the box in the lower left.
}
\label{PolCorA}
\end{figure*}

\begin{figure*}
\centerline{ 
  \includegraphics[height=2.2in]{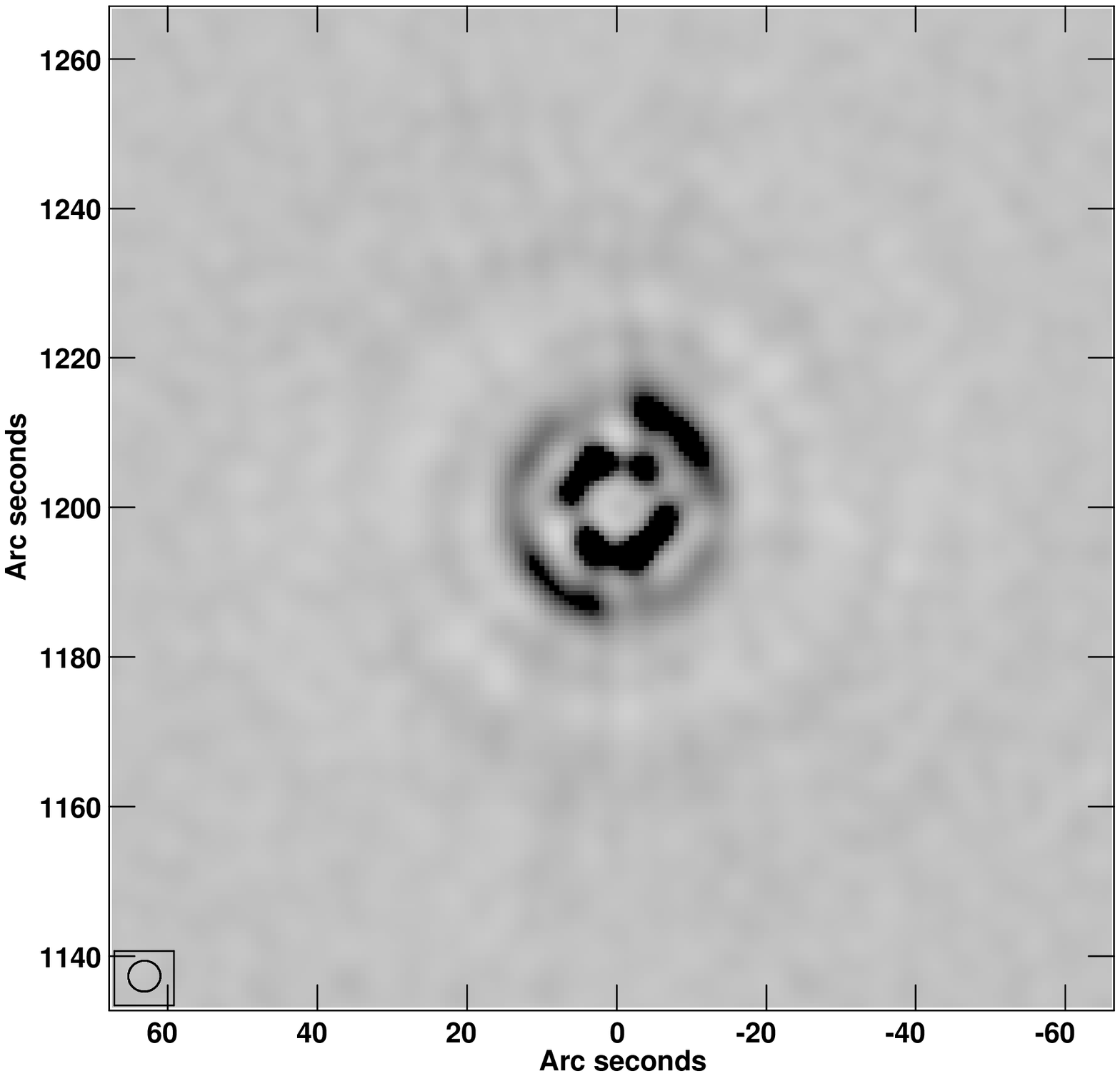}
  \includegraphics[height=2.2in]{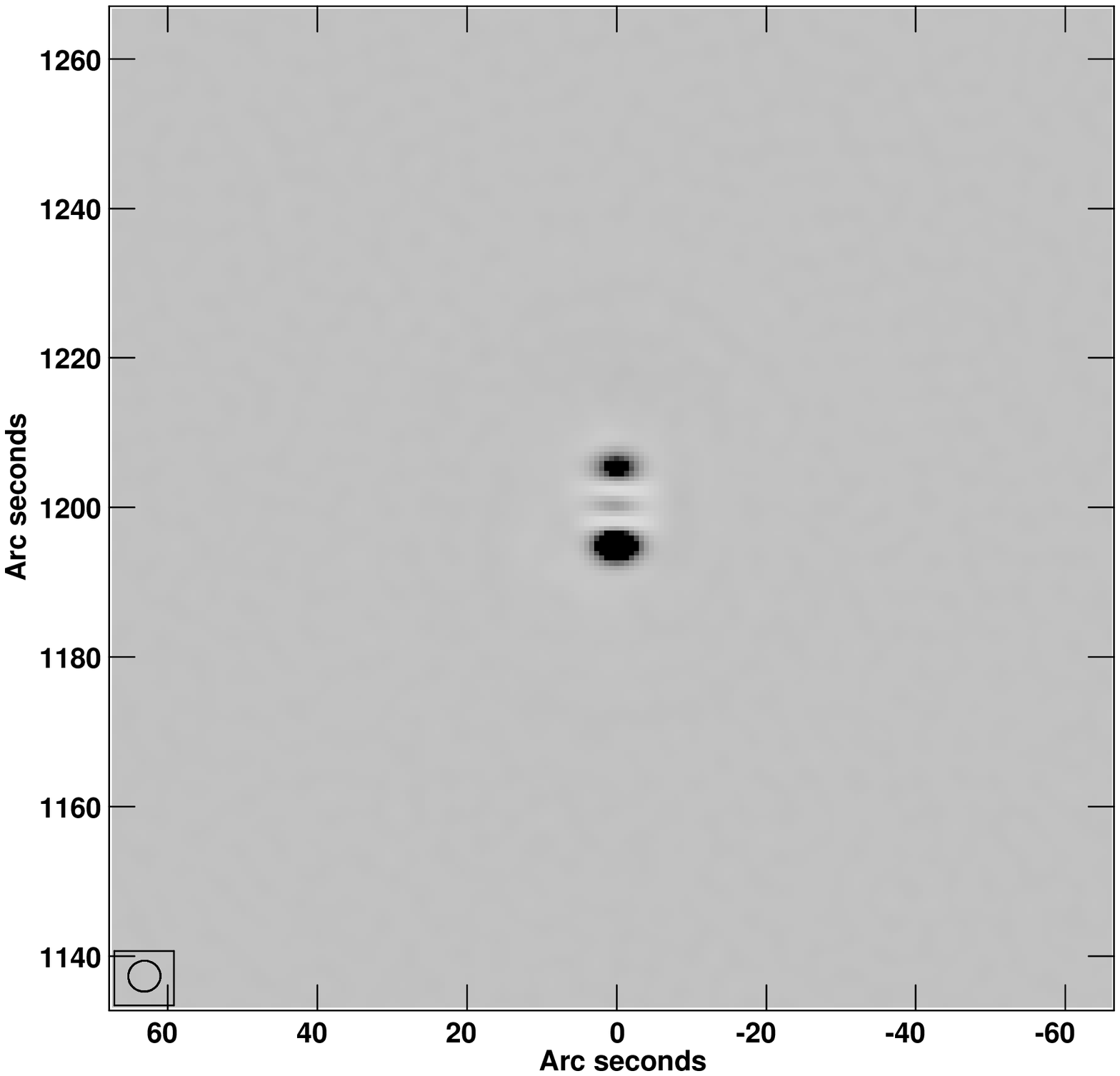}
  \includegraphics[height=2.2in]{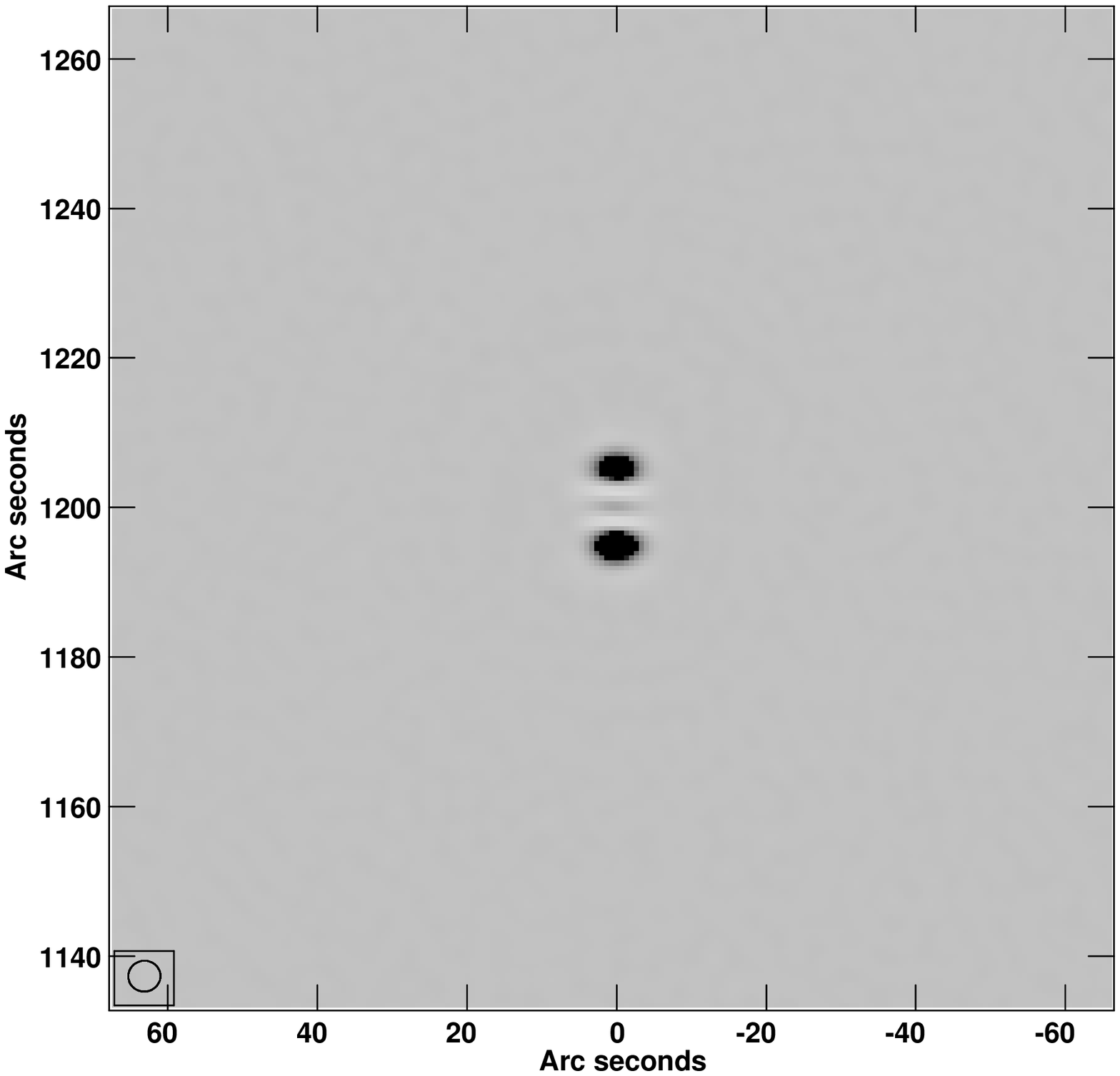}
}
\centerline{ 
  \includegraphics[height=2.2in]{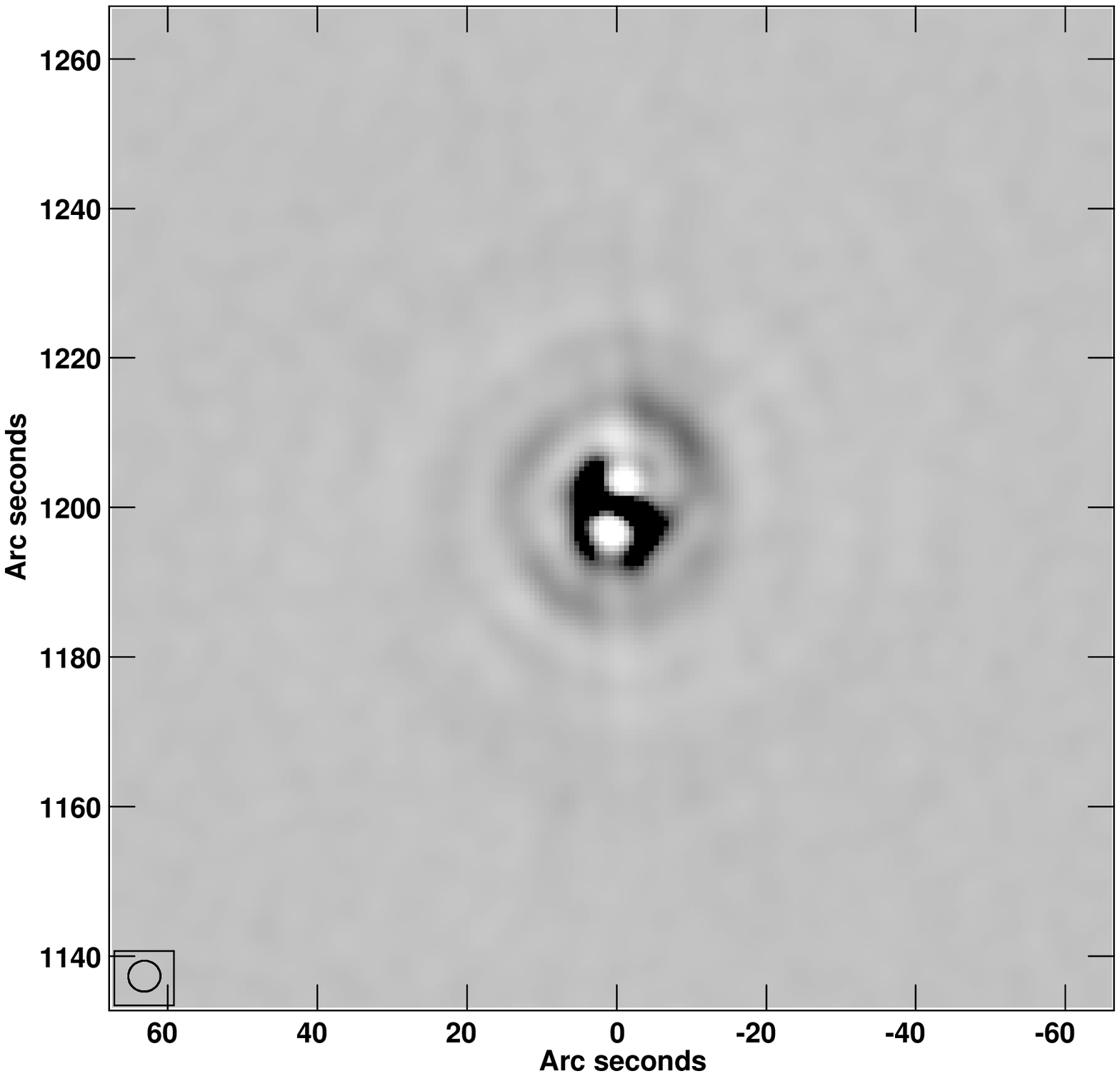}
  \includegraphics[height=2.2in]{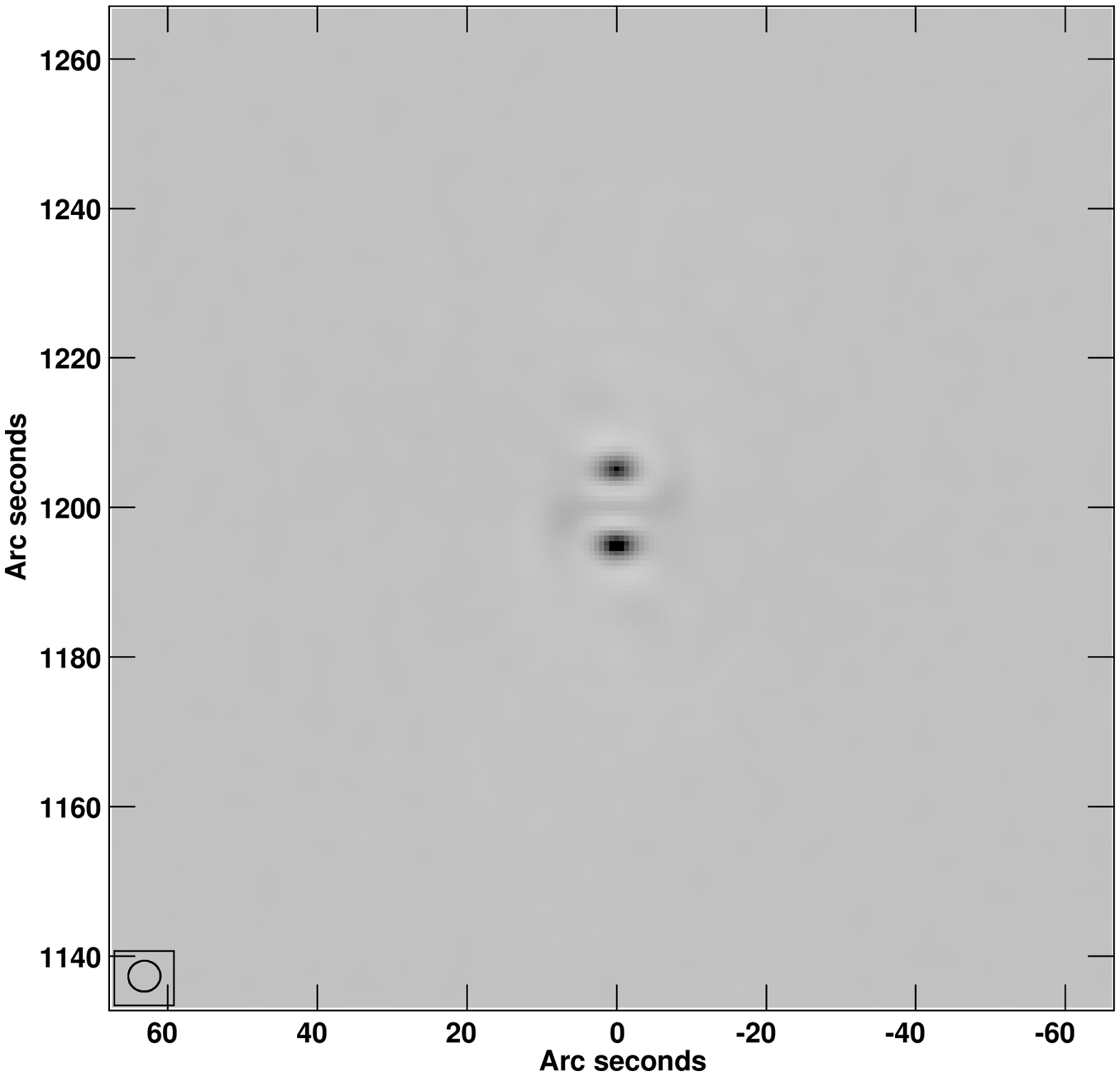}
  \includegraphics[height=2.2in]{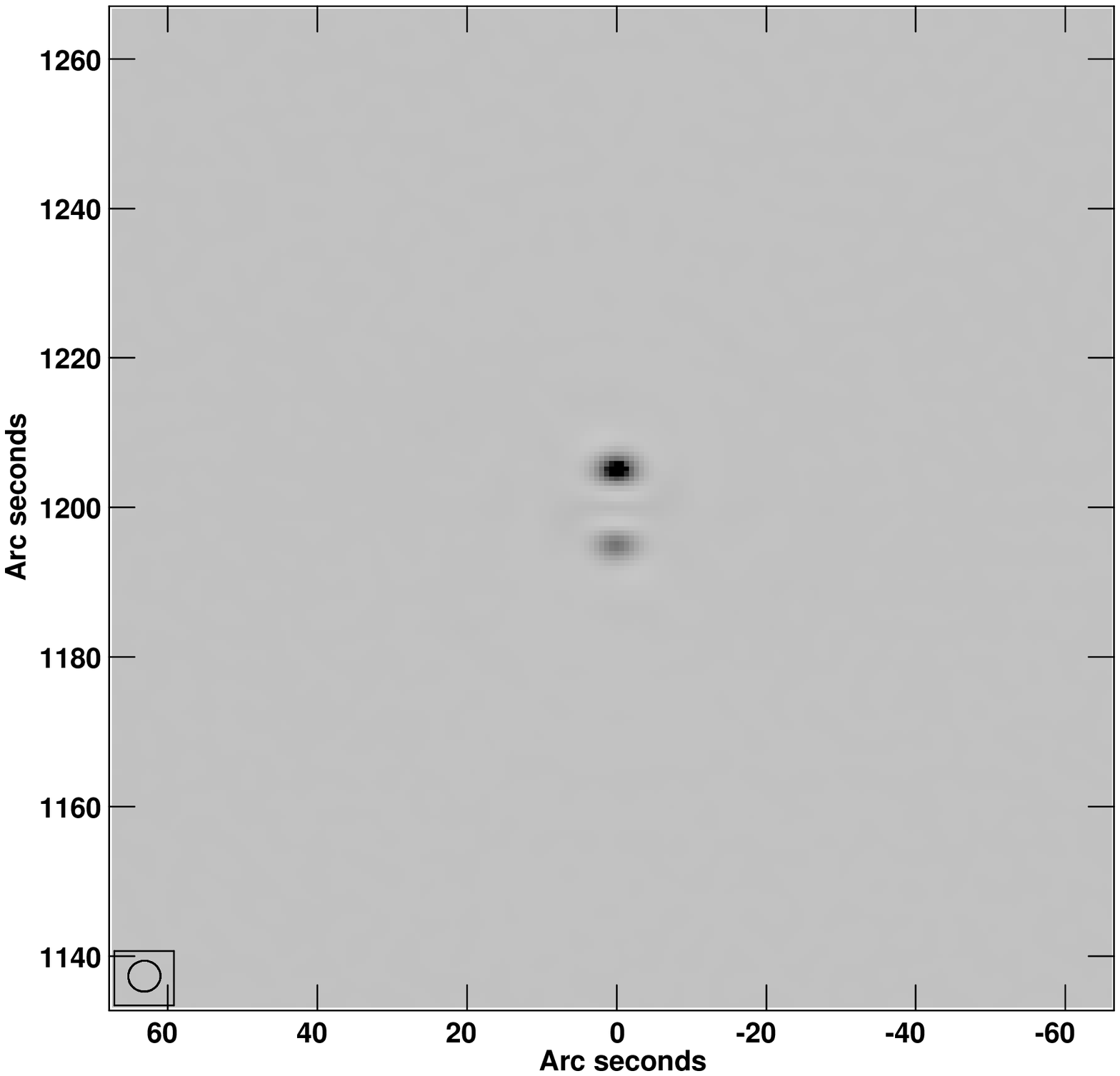}
}
\centerline{ 
  \includegraphics[height=2.2in]{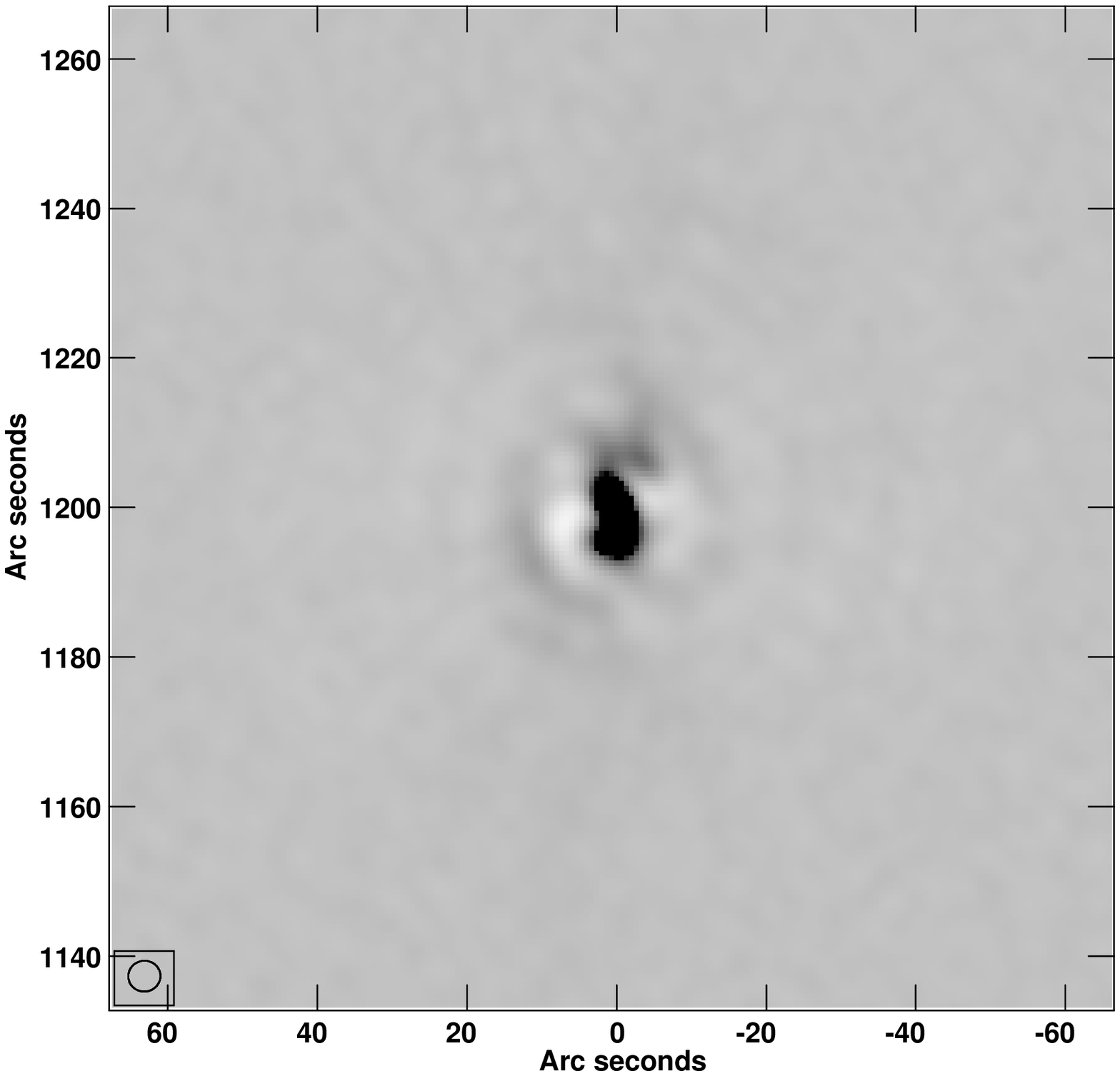}
  \includegraphics[height=2.2in]{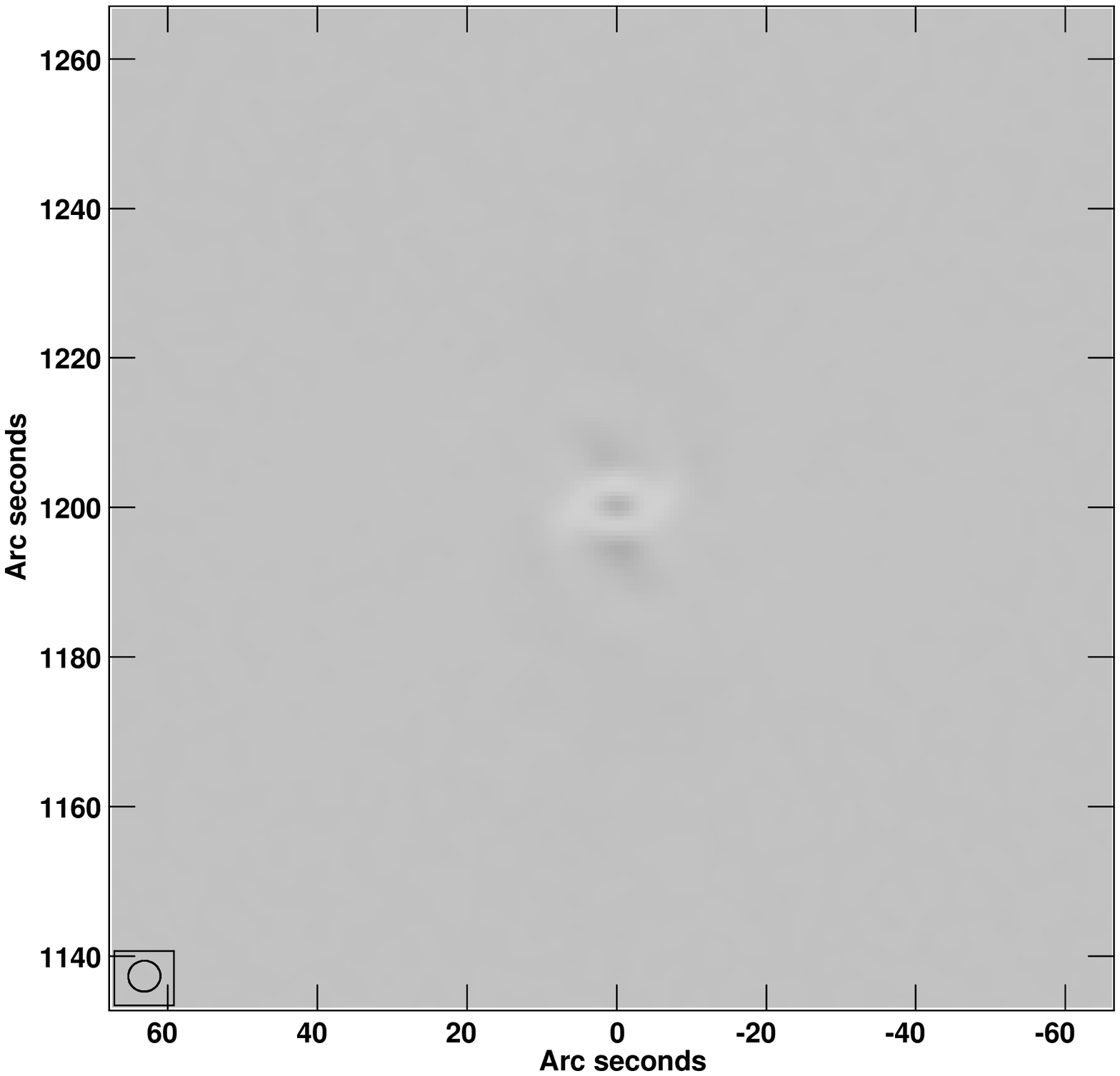}
  \includegraphics[height=2.2in]{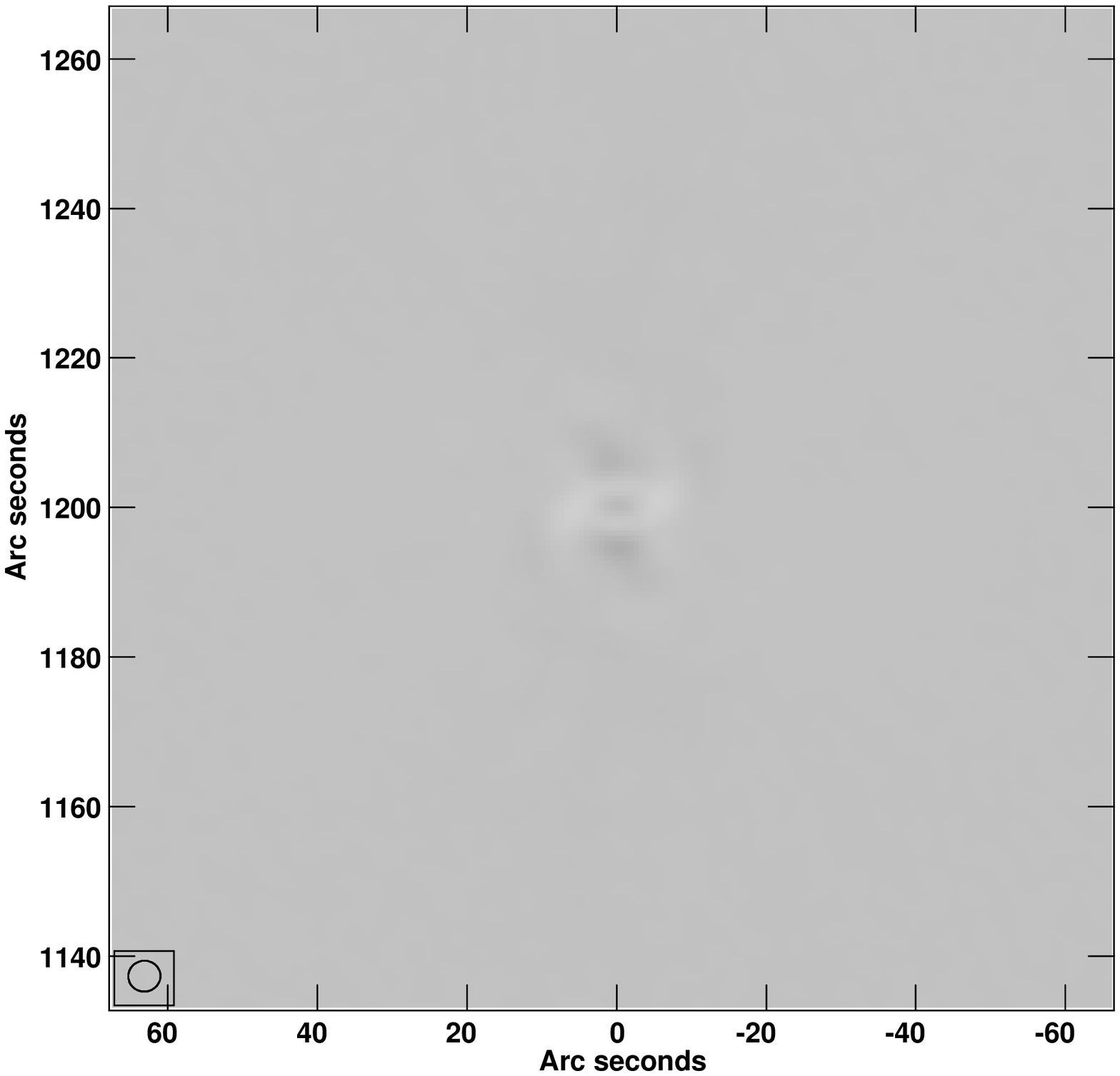}
}
\centerline{ 
  \includegraphics[height=2.2in]{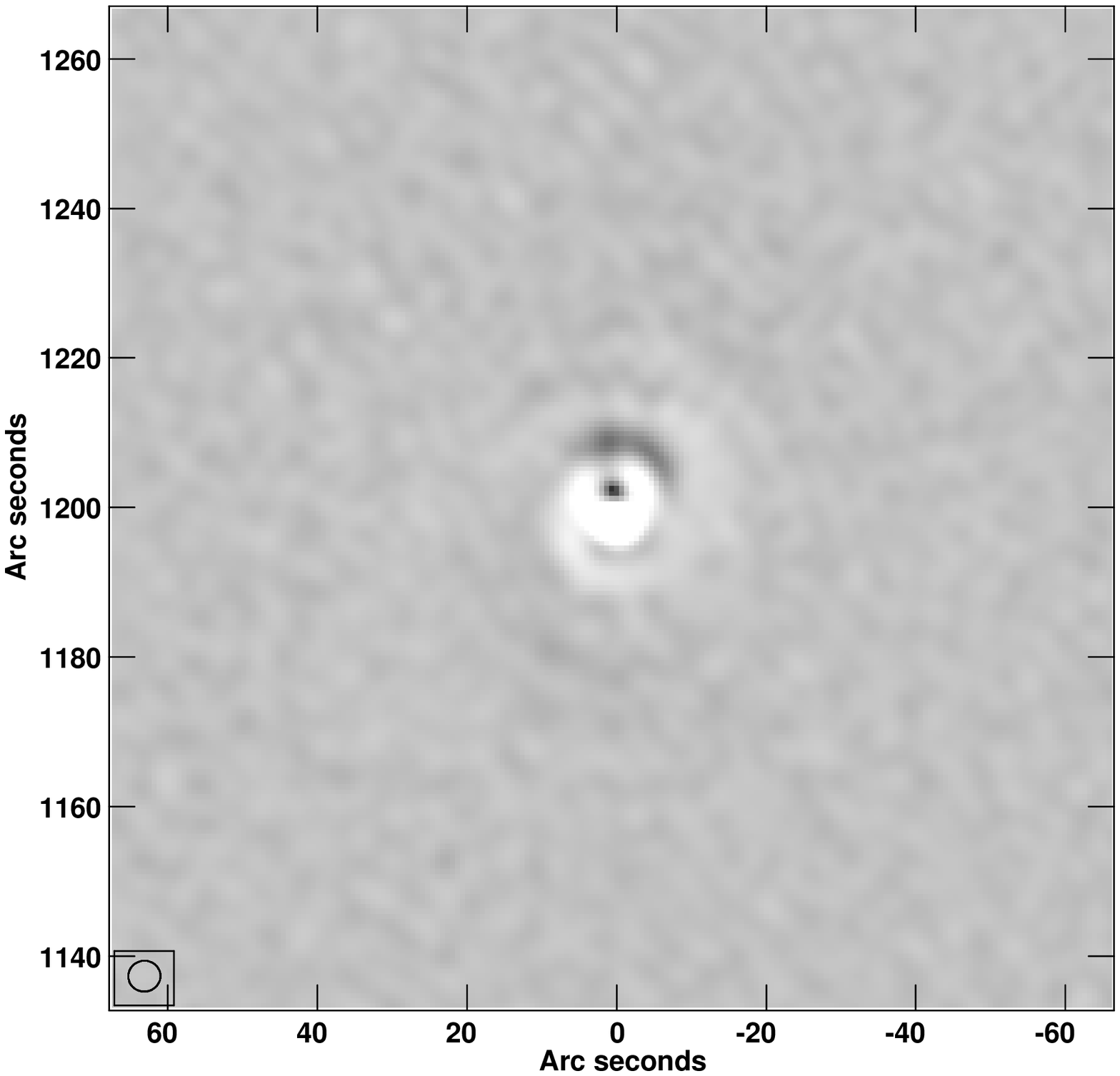}
  \includegraphics[height=2.2in]{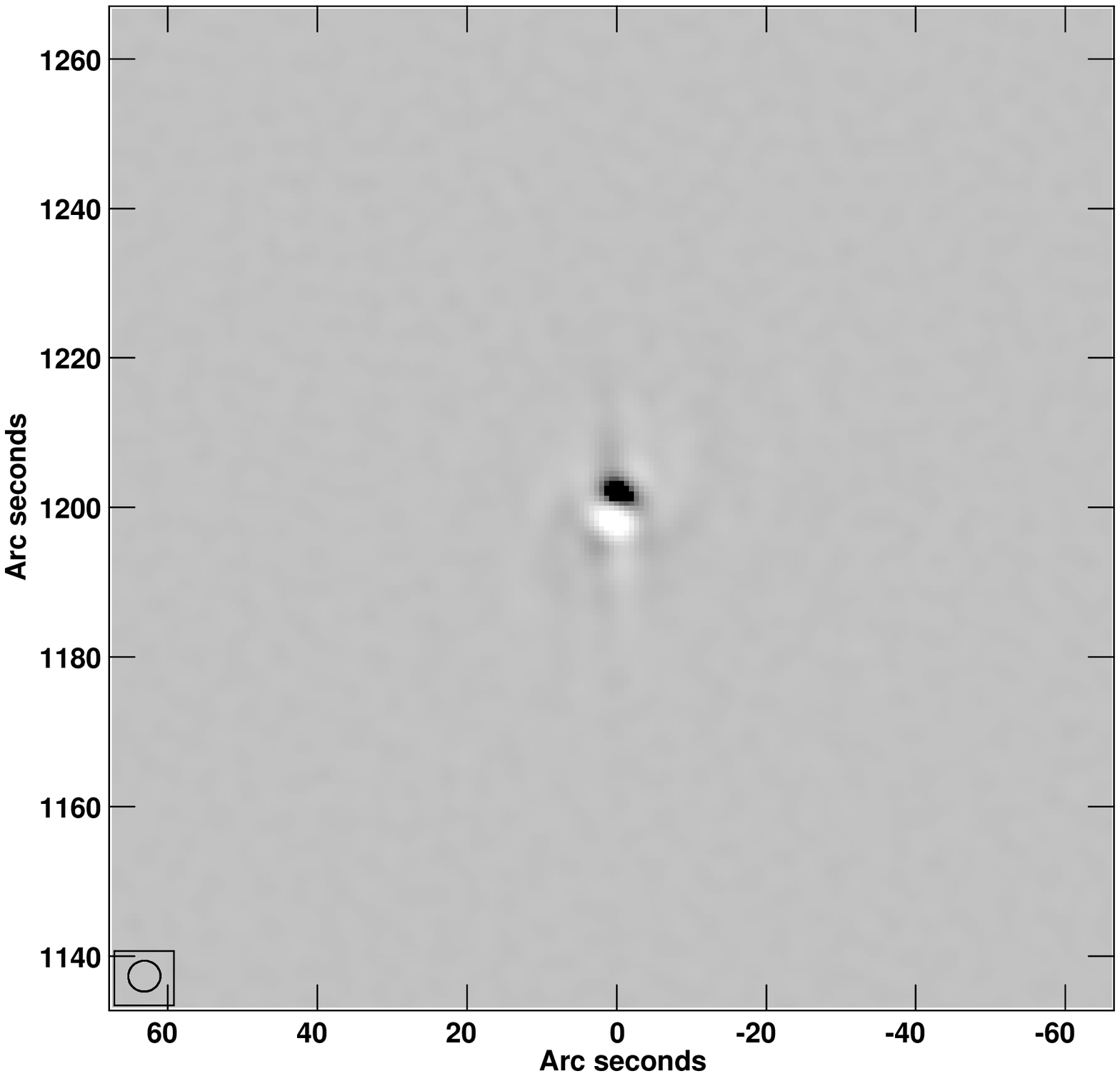}
  \includegraphics[height=2.2in]{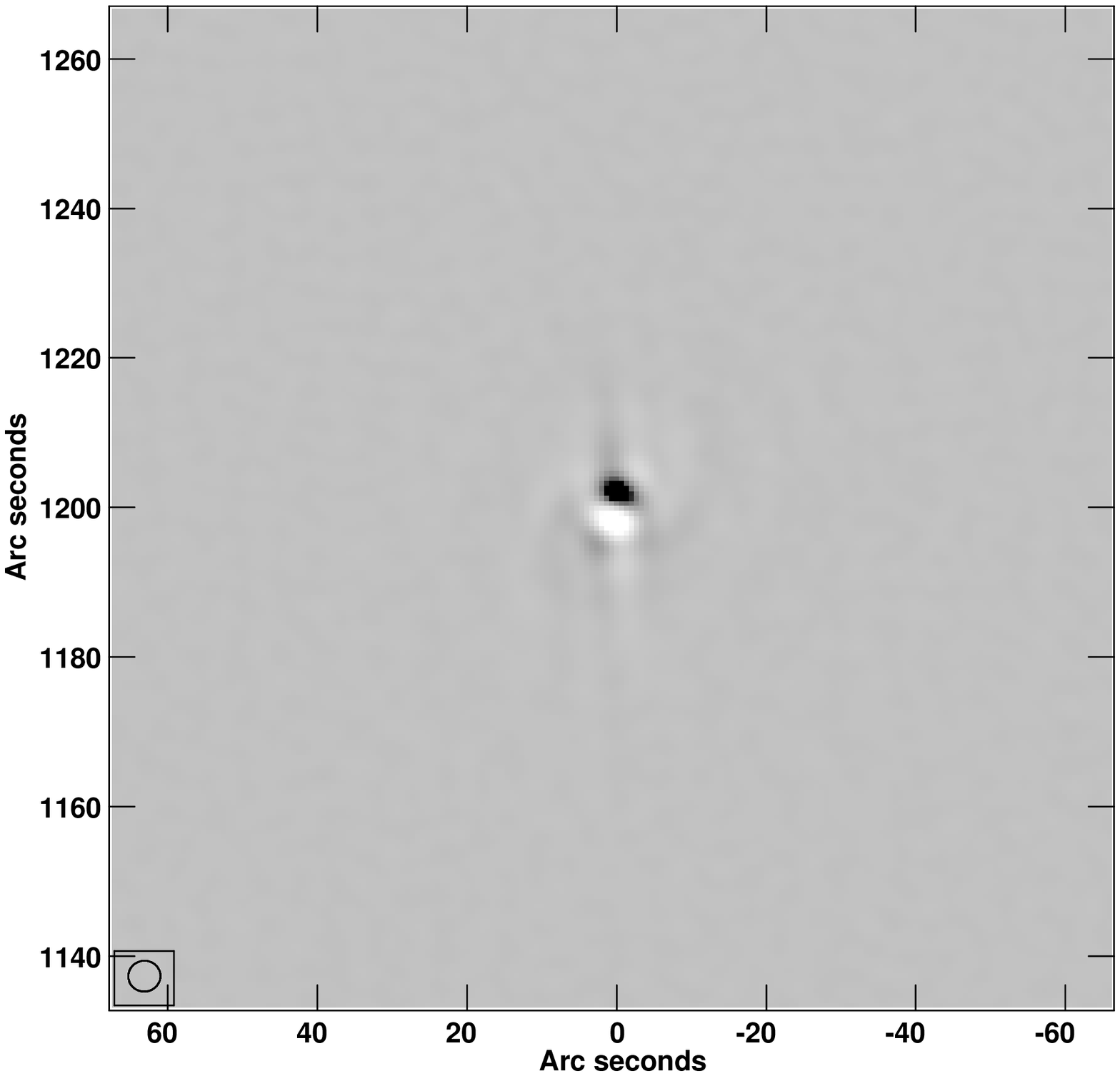}
}
\caption{Like Figure \ref{PolCorA} but source B.
}
\label{PolCorB}
\end{figure*}

\begin{figure}
\centerline{ \includegraphics[width=3.0in,angle=-90]{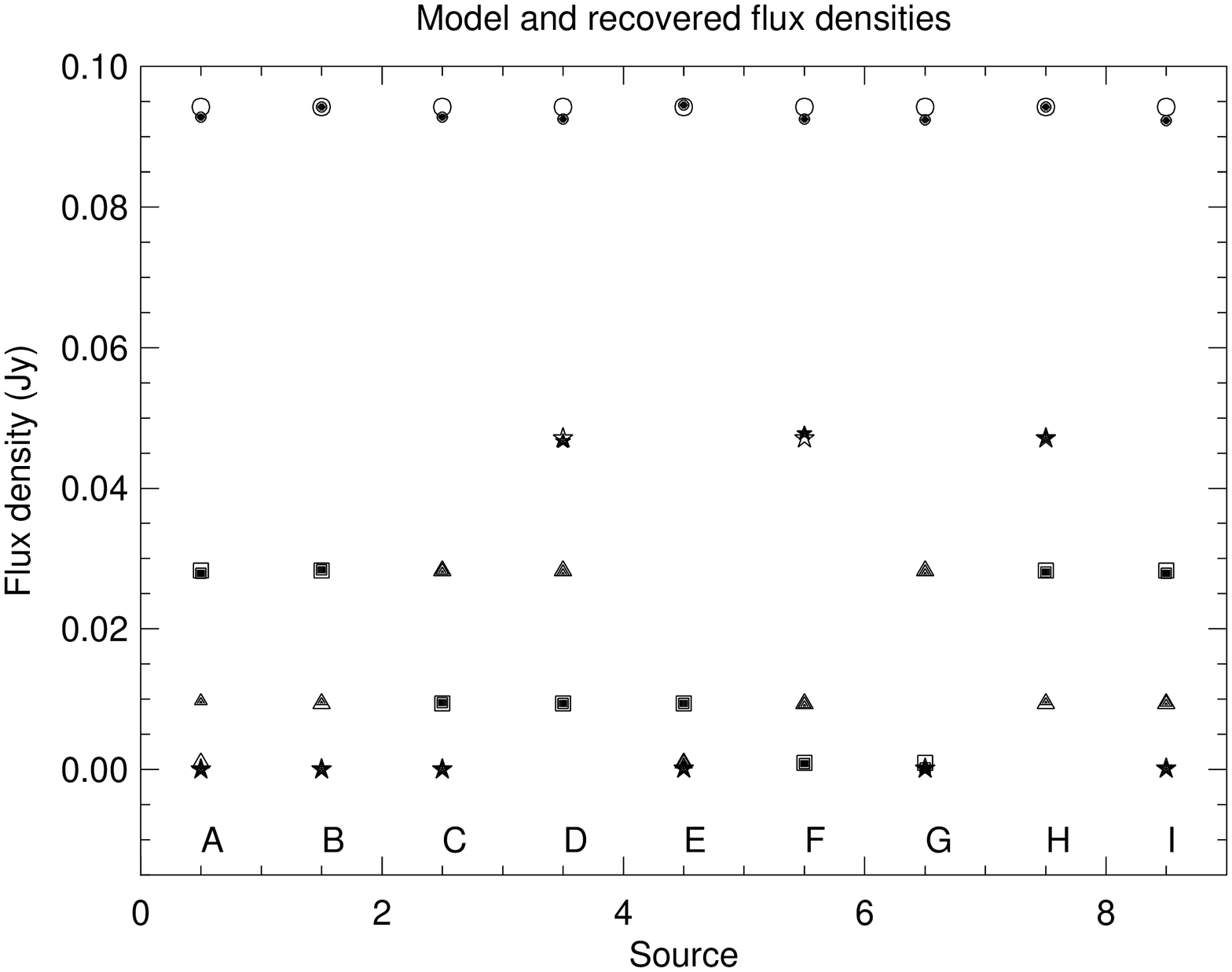}}
\caption{Comparison of simulation model flux densities and those
  recovered from imaging. 
Sources are numbered A--I, open symbols are the input model and filled
symbols are the values recovered from imaging.
Stokes I is a circle, Stokes Q a square, Stokes U a triangle and
Stokes V a star.
}
\label{PolCompare}
\end{figure}

Residual images of these three processings are shown in Figures
\ref{PolCorA}\&\ref{PolCorB} for all four Stokes parameters for two
of the sources.
After MFBeam, the resultant images were primary beam corrected using
the ``perfect'' beam shape and the fitted flux densities are compared
with the model going into the simulation in Figure \ref{PolCompare}.

\section{Discussion}
Asymmetries in, and differences among, antenna patterns in a radio
interferometric array in an extended synthesis result in artifacts
arising from brighter sources which are capable of obscuring fainter
sources. 
These differences in antenna patterns will also result in spurious
polarized responses to sources bright in total intensity (Stokes I).
The preceding sections have explored examples of these and their
corrections. 
The most important features of the process for correcting beam
asymmetries and differences among antenna patterns are 1) how well the
artifacts are reduced and 2) the relative expense of the computations.
\subsection{Accuracy}
The results of the realistic, SKADS, simulation shown in Figure
\ref{SKADSImage} centered on several of the brightest sources which have
very significant levels of artifacts when uncorrected.
These artifacts are reduced to the level of those in images derived
from data without the beam shape corruptions added.
The overall ``noise'' level in the image was reduced by over a factor
of four in this test.
Figures \ref{SKADSFlux} \& \ref{SKADS_SI} show that the flux densities
of sources at all flux density levels were well recovered and the
spectral indices of the sources recovered do not show a significant bias.

Spurious polarization responses to Stokes I generally are very small
at the pointing center and increase away from it.
The test using the grid of unpolarized sources in
Table \ref{GridTab}  and Figure \ref{MaxArt} shows that the corrections
reduce spurious linearly polarized artifacts at the edge of the field
of imaged from $\sim$1.0$\times 10^{-3}$ to $\sim$4.0$\times 10^{-5}$
and spurious circular polarization drops from $\sim$6.0$\times
10^{-4}$ to $\sim$1.0$\times 10^{-5}$. 

The test using the grid of partially polarized sources, Figures
\ref{PolCorA} - \ref{PolCompare}, shows that polarized artifacts were 
reduced to the level of those in the uncorrupted data and that the
polarized flux densities of the input models are well recovered.
This is true for sources with a significant range of fractional
polarizations. 

\subsection{Performance}
The CLEAN deconvolution is a very nonlinear process and slight
differences can cause it to take rather different paths; this makes
comparing the run times of tests with different algorithms tricky.
The realistic sky simulation described in section \ref{SKADS} gives
the best estimate of the increased cost of the beam correction
described here.
Imaging the corrupted data with beam corrections took 50\% longer
(2.96 v 2.00 hours) than without.

This test was performed using a CPU implementation of the model
calculations as a GPU implementation is not yet available for the model
with beam corrections and the machine used for the test did not have a
GPU. 
GPU implementation of the sky model calculation dramatically changes
the economics of imaging such that it usually becomes a minor
component of the overall imaging computation.

The interferometer response to a given sky model does need to be
accurately calculated, perhaps more so when the corrupting effects are
not corrected in the imaging.
In the technique described here, this is done via a ``DFT'' model
calculation requiring large numbers of sine/cosine calculations.
While these computations dominate the floating point operations of the
imaging process, they are highly parallelizable; there are no
dependencies which allows very efficient use of vector (SSE/AVX),
multithreading and especially GPU implementations.
GPU implementations of the DFT model calculation can reduce their cost
to a small fraction of the total run time.

\subsection{Comparison with other Techniques}
The principle difference between the technique described here and
previous methods of correction of beam effects described in Section
\ref{previous} is that no attempt is made to correct the effects in
the image formation.
Instead, the calculation of the response to the sky model subtracted
from the visibility data in each major CLEAN cycle includes these beam
effects. 
As the process proceeds, the visibility/image residuals approach zero
and the accumulated sky model approaches the desired one.

In the regime that the image corruptions are relatively minor,
amplitude only ones with only small distortions, and no shift of
source positions, the correction of dirty/residual images formed
during deconvolution appears unnecessary. 
We note that effects such as those from the ionosphere which are
largely in phase can, and do, introduce time dependent shifts in
apparent position.
These effects may seriously distort sources and thus must
be corrected in the image formation. 
Techniques for correcting dirty/residual images such as ``A
Projection'' can become quite expensive and yet still require
corrections to be made in calculating the interferometer response to
the sky model. 
The expense appears unwarranted in at least an interesting range of
use cases.  

``Degridding'' of Fourier transforms of sky models is frequently done
by interpolation of 2D complex grids at the spatial frequencies of the
visibility measurements.
The ``DFT'' method described here allows a much more accurate
calculation of the instrumental response for each visibility
measurement than interpolation from a grid.
A relatively arbitrary model of the instrumental response can be
calculated. 
While the DFT modeling requires a substantially larger number of
floating point operations, the lack of dependencies allows efficient
usage of modern computing hardware (SIMD instructions, threading, GPU)
largely eliminates the runtime differences.

Faceted imaging has a computational advantage over full field
techniques such as W stacking or W projection in that many facets can
be gridded and imaged in parallel using the same input data.
Furthermore, once the full field has been imaged, only facets with
emission likely to be CLEANed need be imaged in a given major cycle.

\section{Summary\label{Summary}}
At low frequencies and with small antennas, the field of view of radio
interferometers is quite large and with the sensitivity of modern
arrays, sources can be detected out into the side-lobes of the antenna
pattern where there are usually significant asymmetries in the antenna
gain. 
For 2 axis antennas with alt-az mounts this pattern rotates on the sky
causing  apparent brightness variations in the sources.
This causes artifacts in the images which are non convolutional and,
if uncorrected, can limit the dynamic range of the image.
Heterogeneous arrays are an even more extreme case.

Furthermore, the spurious polarized instrumental response of the
antennas increases with radius from the pointing direction and is
also asymmetric.
The spurious polarized response to Stokes I may corrupt, or even
swamp, true polarized emission.
Rotation of the pattern with parallactic angle will tend to reduce the
instrumental polarization but generally not eliminate it.

A technique for making wide area beam full polarization corrections is
described and shown to be effective for full polarization imaging when
used in a CLEAN--like deconvolution.
The general technique is to ignore beam effects when making
dirty/residual images but then using an accurate beam model in
computing the instrumental response to a partial sky model to be
subtracted from the residual data before computing the next residual
image.
After multiple major cycles, the residual data converges towards zero
and the accumulated sky model converges towards the correct one.

For wide area imaging a critical part of the technique is to 
correct the instrumental response  from the actual, poorly behaved
one(s), to a single well behaved, ``perfect'', i.e. real, symmetric, one.
This avoids amplifying the noise in parts of the image in which the
primary beam gain is low but which contains strong sources needing to
be included in the deconvolution.
For an array using multiple antenna designs with multiple diameter
antennas, using the ``perfect'' beam of the smaller antenna seems to
give better results.

A number of tests using simulated data show that the technique is
effective and relatively efficient; with GPU implementation the
additional cost of the sky model calculation can be reduced to a minor
component of the imaging cost.
A test of a realistic sky shows that high dynamic range, full field
imaging is possible even with the heterogeneous array used for the
imaging. 
The source flux densities and spectral indices of the input model are
well recovered (Figures \ref{SKADSFlux} \& \ref{SKADS_SI}).

Spurious polarized responses due to beam effects were greatly reduced
as is shown in Table \ref{GridTab} and Figure \ref{MaxArt}.
Beam artifacts around polarized sources were reduced to the level of
the basic imaging, see Figures \ref{PolCorA} \& \ref{PolCorB} and
model polarized flux densities are well recovered, see Figure
\ref{PolCompare}. 

The tests described here are for the MeerKAT+ array under development
which includes two antenna designs, 1) MeerKAT 13.5 m dishes
and 2) SKA design 15 m antennas.
With the field of view and dynamic range requirements of this array
and the very different antenna patterns, beam corrections will be
needed all, or nearly all, of the time.
This is an extreme case but a solution that works for it should also
work for a homogeneous array with asymmetric antennas.

\section*{Acknowledgments}
The MeerKAT telescope is operated by the South African Radio Astronomy
Observatory, which is a facility of the National Research Foundation,
an agency of the Department of Science and Innovation.  
The National Radio Astronomy Observatory is a facility of the National
Science Foundation, operated under a cooperative agreement by Associated
Universities, Inc.
We would like to thank the anonymous referee for encouraging a
discussion in a larger context, leading to an improvement in the
document. 

\vspace{5mm}
\facilities{MeerKAT}

\software{Obit \cite{OBIT}}
\bibliography{BeamCor}{}
\bibliographystyle{aasjournal}


\end{document}